\documentclass[vecphys]{svmult}

\usepackage{makeidx}     % allows index generation
\usepackage{graphicx}    % standard LaTeX graphics tool
\usepackage{multicol}    % used for the two-column index
\usepackage[bottom]{footmisc}% places footnotes at page bottom
\usepackage{bm}
\makeindex             % used for the subject index
\begin{document}

\frontmatter%%%%%%%%%%%%%%%%%%%%%%%%%%%%%%%%%%%%%%%%%%%%%%%%%%%%%%

%\include{dedic}
%\include{pref}

%\tableofcontents

%\include{cblist}

\mainmatter%%%%%%%%%%%%%%%%%%%%%%%%%%%%%%%%%%%%%%%%%%%%%%%%%%%%%%%
% %%%%%%%%%%%%%%%%%%%%%%%%%% author.tex %%%%%%%%%%%%%%%%%%%%%%%%%
% %
% % sample root file for your contribution to a "contributed book"
% %
% % "contributed book"
% %
% % Use this file as a template for your own input.
% %
% %%%%%%%%%%%%%%%%%%%%%%%% Springer-Verlag %%%%%%%%%%%%%%%%%%%%%%%%%%

% % RECOMMENDED %%%%%%%%%%%%%%%%%%%%%%%%%%%%%%%%%%%%%%%%%%%%%%%%%%%
% \documentclass[multphys,vecphys,sprmidx]{svmult}
% %\documentclass[10pt]{article}
% % choose options for [] as required from the list
% % in the Reference Guide, Sect. 2.2
% %\bibliographystyle{unsrt}
% \usepackage{makeidx}         % allows index generation
% \usepackage{graphicx}        % standard LaTeX graphics tool
%                              % when including figure files
% \usepackage{amssymb}
% \usepackage{bm}
% \usepackage{multicol}        % used for the two-column index
% \usepackage[bottom]{footmisc}% places footnotes at page bottom
% %\usepackage{epstopdf}
\DeclareGraphicsRule{.tif}{png}{.png}{`convert #1 `dirname #1`/`basename #1 .tif
`.png}
% etc.
% see the list of further useful packages
% in the Reference Guide, Sects. 2.3, 3.1-3.3

%\makeindex            % used for the subject index
                       % please use the style sprmidx.sty with
                       % your makeindex program

\def\beq{\begin{equation}}
\def\eeq{\end{equation}}
\def\beqa{\begin{eqnarray}}
\def\eeqa{\end{eqnarray}}

\newcommand{\ds}{\displaystyle}
\newcommand{\bfr}{{\bf r}}
\newcommand{\bfg}{{\bf g}}
\newcommand{\bfR}{{\bf R}}
\newcommand{\bfv}{{\bf v}}
\newcommand{\bfF}{{\bf F}}
\newcommand{\bff}{{\bf f}}
\newcommand{\bfJ}{{\bf J}}
\newcommand{\bfk}{{\bf k}}
\newcommand{\bfp}{{\bf p}}
\newcommand{\bnabla}{{\bm \nabla}}
\newcommand{\bxi}{{\bm \xi}}
\newcommand{\btau}{{\bm \tau}}
\newcommand{\bsigma}{{\bm \sigma}}
\newcommand{\bomega}{{\bm \omega}}
\newcommand{\bOmega}{{\bm \Omega}}
\newcommand{\bGamma}{{\bm \Gamma}}
\newcommand{\blue}{\color{blue}}
\newcommand{\uvec}{\hat{\bf u}}
\newcommand{\nvec}{\hat{\bf n}}
\newcommand{\la}{\langle}
\newcommand{\tm}{\tilde{m}}
\newcommand{\ran}{\rangle_\uvec}
\newcommand{\rantwo}{\rangle_{\uvec_2}}
\newcommand{\exangle}{|\uvec_1\times\uvec_2|}
\newcommand{\trho}{\tilde{\rho}}
\newcommand{\tsigma}{\tilde{\sigma}}
\newcommand{\nhat}{\hat{n}}
\newcommand{\uhat}{\hat{u}}
\newcommand{\chat}{\hat{c}}
\newcommand{\tgp}{\tilde{\gamma}_P}
\newcommand{\tgnp}{\tilde{\gamma}_{NP}}

%%%%%%%%%%%%%%%%%%%%%%%%%%%%%%%%%%%%%%%%%%%%%%%%%%%%%%%%%%%%%%%%%%%%%

%\begin{document}

%\input{notes}

%\shownotes

\title*{Hydrodynamics and rheology of active polar filaments}
%\title*{Living soft matter : hydrodynamics and rheology of solutions
%of active polar filaments}
\author{Tanniemola B. Liverpool\inst{1}\and
M. Cristina Marchetti\inst{2}}
\institute{Department of Applied Mathematics, University of Leeds,
Woodhouse Lane, Leeds LS2 9JT, UK \texttt{t.b.liverpool@leeds.ac.uk} \and
Physics Department, Syracuse University, Syracuse, NY 13244, USA
\texttt{mcm@phy.syr.edu}}
\maketitle

\abstract{The cytoskeleton provides eukaryotic cells with mechanical
  support and helps them perform their biological functions.  It is a
  network of semiflexible polar protein filaments and many accessory
  proteins that bind to these filaments, regulate their assembly, link
  them to organelles and continuously remodel the network.  Here we
  review recent theoretical work that aims to describe the
  cytoskeleton as a polar continuum driven out of equilibrium by
  internal chemical reactions.  This work uses methods from soft
  condensed matter physics and has led to the formulation of a general
  framework for the description of the structure and rheology of
  active suspension of polar filaments and molecular motors.}

\section{Introduction}
\label{sec:m1}
% Always give a unique label
% and use \ref{<label>} for cross-references
% and \cite{<label>} for bibliographic references
% use \sectionmark{}
% to alter or adjust the section heading in the running head

Cells are living soft matter. They are composed of a variety of soft
materials, such as lipid membranes, polymers and colloidal aggregates,
often constrained to a reduced spatial dimensionality and geometry. It
is then reasonable to expect that the dynamics and interactions of
these constituents that control cell function takes place on the same
time and energy scales as those of synthetic soft materials. Life
adds, however, a new feature not found in traditional soft matter: the
constant flow of energy and information required to keep living
organisms alive. This new feature makes cells of particular interest
to physicists as understanding the behavior of {\em active} living
matter requires the development of new theoretical concepts and
experimental techniques.

The eukaryotic cell cytoskeleton is a perfect example of this novel
type of {\em active material}.  The cytoskeleton allows the cell to
carry out coordinated and directed movements such as cell crawling,
muscle contraction, and all the changes in cell shape in the
developing embryo~\cite{Alberts_m}. The cytoskeleton also supports
intra-cellular movements such as the transport of organelles in the
cytoplasm and segregation of chromosomes during cell
division~\cite{howardBook}.
%The cytoskeleton contains a large
%number of different proteins some of which are characterized and
%many of which remain unknown with their functions yet to be
%identified.
It is highly inhomogeneous, with a large variety of different
dynamical {\em supramolecular} structures. Examples are contractile
elements like stress fibres, or the contractile ring in mitosis, or
astral objects like the mitotic spindle which forms during cell
division~\cite{Alberts_m,howardBook}.

Self-assembled filamentous protein aggregates play an important role
in the mechanics and self-organization of the cytoskeleton.  In
addition, a number of other proteins interact with them and modulate
their structure and dynamics.  Cross-linking proteins bind to two or
more filaments together to form a dynamical gel.  Molecular motor
proteins bind to filaments and hydrolyze nucleotide Adenosine
triphosphate (ATP). This process, coupled to a corresponding
conformational change of the protein, turns stored chemical energy
into mechanical work. Capping proteins modulate the polymerization and
depolymerization of the filaments at their ends.

A key question is how the elements of the cytoskeleton cooperate to
achieve its function. To what extent is there a 'cellular' brain and
how closely does it control cellular mechanisms?  How much of a role
does spontaneous self-organization driven by general physical
principles play?

%Can the constraints imposed by the laws of physics
%guide us in our search for possible strategies used to achieve
%cellular function.

Much of the recent progress in the understanding of the complex
structures and processes that control the behavior and function of the
cytoskeleton has been linked to the development of new biophysical
probes allowing an unprecedented view of sub-cellular processes at
work. Mechanical probes such as optical and magnetic tweezers
\cite{mehta98}, atomic force microscopes \cite{Dammer} and
micropipettes probe the response of the elements of the cytoskeleton
to locally applied forces. Visualization techniques using fluorescence
microscopy, e.g. fluorescence imaging with one-nanometer accuracy
\cite{fiona} or single-molecule high-resolution co-localization
\cite{shrec} based on organic dyes
%and more recently quantum dots,
allow one to follow the dynamics of single molecules inside living
cells, ({\em in-vivo})
%% within the organism
giving insights into the microscopic processes underlying cellular
dynamics. Many of these experimental developments are reviewed in this
book.
%Flourescence Recovery After Photobleaching FRAP
%Fluorescence Resonance Energy Transfer FRET
%Fluorescence Correlation Spectroscopy FCS
%fluorescence imaging with one-nanometre accuracy FIONA
%single-molecule high-resolution colocalization SHREC

Because of the large number of unknown components, it is also of
interest to study simplified systems consisting of a smaller number of
well characterized elements, {\em in-vitro}.  This has led to a number
of experimental biophysical studies of purified solutions of
cytoskeletal filaments and associated molecular motors that have
established that motor-induced activity drives the formation of a
variety of spatially inhomogeneous patterns, such as bundles, asters
and
vortices~\cite{Szent-Gyorgi51,Trinick79,takiguchi,urrutia,nedelec97,nedelec2,surrey01,Backouche06}.
These are reminiscent of some of the supramolecular structures present
in the cytoskeleton~\cite{waterman,verkhovsky}. The mechanical
properties of filament-motor systems has also been studied showing
qualitative differences from passive filament suspension. Because of
the controlled nature of their preparation and the detailed knowledge
of their constituents, {\em in vitro} studies are particularly
amenable to a quantitative description using techniques from
theoretical physics. In this review we will be mostly concerned with
describing the behavior of such simplified systems on large time
scales (times $\ge$microsecond) where the atomistic details are not
important and a coarse-grained phenomenological description may
suffice.

The reductionist viewpoint typified by this approach also has its
drawbacks.  A simplified system necessarily can provide only a subset
of the phenomena observed in living cells since only a small fraction
of the components are present.  A choice must also be made of {\em
  which} simplified system to study as different combinations of
components may give different or similar behavior.  This choice must
of course be heavily influenced by previous
experiments~\cite{Szent-Gyorgi51,Trinick79}. A living cell is a highly
optimized complex {\em system} of interacting agents with the ability
to modulate its response to complex changes in its environment. This
complexity will be missing from simple mechanical models described
here.  There is some hope, however, that this complexity can
eventually be combined with the physical picture emerging from the
approach we present here to give a more complete "biophysical''
picture of motility in cells in which the laws of physics provide
important constraints on the possible "system" dynamics. Finally, even
within our limited frame of reference, we will also make a number of
simplifying assumptions in developing the models. Some of the
important physical phenomena ignored here, such as active
polymerization, treadmilling~\cite{dogterom95,mogilner03} and filament
flexibility~\cite{isamag,kasmacjan,kroyfrey,morse98,freds98,everaers,storm05},
will be incorporated in future work.

%In the next section, we introduce and describe two families of
%cytoskeletal filaments and their associated motors. In Section
%\ref{sec:experiments}, we describe some recent in-vivo and
%in-vitro experiments tracking the motion and mechanics of these
%filaments and motors.
%We also describe recent in-vitro experiments on purified
%filament/motor mixtures.
We first review some recent theoretical approaches to describe active
filament suspensions. We then describe some of our current work and
give perspectives for the future.

\section{Theoretical modeling of active systems} \label{sec:m2} There
have been a number of recent theoretical studies of the collective
dynamics of mixtures of rigid filaments and motor clusters. First and
most microscopic, numerical simulations with detailed modeling of the
filament-motor coupling have yielded patterns similar to those found
in experiments~\cite{nedelec97,surrey01}, including vortices and
asters. These simulations modeled the filaments as elastic rods with
motor clusters being parameterized by three binding parameters, the on
and off rates and the off-rate at the plus end of the filament. It was
found that the rate of motor unbinding at the polar end of the
filaments plays a crucial role in controlling the vortex to aster
transitions at high motor densities~\cite{nedelec2}.

A second interesting development has been the proposal of 'mesoscopic'
mean-field kinetic equations first studied in one dimension
\cite{nakazawa96,nakazawa98}, where the effect of motors is
incorporated via a motor-induced relative velocity of pairs of
filaments, with the form of such a velocity inferred from general
symmetry considerations. Kruse and
collaborators~\cite{kruse00,kruse03,kruse01} proposed a one
dimensional model of filament dynamics and showed the existence of
instabilities from the homogeneous state to contractile
states~\cite{kruse00} and traveling wave solutions~\cite{kruse01}.
We generalized the kinetic model to higher
dimensions~\cite{tblmcm03,tblmcm04} and used it to classify the nature
of the homogeneous states and their
stability~\cite{Ahmadi05,AAMCMTBL06}. Related kinetic models have also
been discussed by other
authors~\cite{ziebert04,ZZ04,Aranson,Aranson06}.

Finally, phenomenological continuum theories have been proposed where
the mixture is described in terms of a few coarse-grained fields whose
dynamics is inferred from symmetry considerations
\cite{bassetti00,lee01,kim03,
  Sankararaman04,kruse04,kruseEPJ06,Voituriez05,Voituriez06,Simha02,Hatwalne04,Muhuri06}.
%Starting from a
%two-dimensional spin lattice model as a microscopic picture,
%theoretical studies have been performed of a sliding assay with
%filaments moving on a bed of motors adsorbed on a flat
%surface.~\cite{bassetti00} Using a combination of mean-field and
%simulation this work obtained a non-equilibrium transitions to
%inhomogeneous states such as stripe patterns.

Lee and Kardar~\cite{lee01} proposed a simple hydrodynamic model for
the coupled dynamics of a coarse-grained filament orientation and the
motor concentration, ignoring fluctuations in the filament density.
These authors argued that filament growth by polymerization provides a
mechanism for an instability of the system from an isotropic to an
oriented state ~\cite{lee01,kim03}, with large-scale aster and vortex
structures. They obtained a phase diagram for the system showing a
transition from vortices to asters. This model was subsequently
generalized by Sankararaman et al.~\cite{Sankararaman04} to include
varying populations of bound and free motors, as well as an additional
coupling of filament orientation to motor gradients.  The effects of
boundary conditions on the steady states of the system was also
studied numerically.

A phenomenological hydrodynamic description for polar gels and
suspensions including momentum conservation has been discussed by
several authors~\cite{kruse04,kruseEPJ06,Voituriez05,Voituriez06}.
These equations generally consider incompressible suspensions and
incorporate momentum conservation in the Stokes approximation, by
assuming the form of constitutive equation for the suspension's stress
tensor on the basis of symmetry consideration. The coupling of flow
and polar order is described via an equation for the local
polarization of the suspension. This model has been used to identify
the non-equilibrium defect structures that can occur in the polar
state~\cite{kruse04,kruseEPJ06} and to analyze the behavior of an
active polar suspension in specific
geometries~\cite{zumdieck05,Voituriez05}. In particular, it was shown
that the interplay between order and activity can yield a spontaneous
flowing state for a solution near a wall~\cite{Voituriez05}. Closely
related hydrodynamic models have been used to describe generically the
collective dynamics of self-propelled particles in solution, such as
swimming bacterial colonies, in both nematic and polar
states~\cite{Simha02,Hatwalne04,Muhuri06}. This work builds on earlier
work by Toner and Tu on hydrodynamic models of flocking, where it was
shown that the nonequilibrium nature of internally driven systems
allows for novel symmetry breaking phase transitions that are
forbidden in equilibrium systems with continuous symmetry in one and
two dimensions~\cite{TonerTu95,TonerTu98,Toner05}.

%%TBL
The main objective of our work has been to establish the connection
between microscopic single-polymer dynamics and the phenomenological
hydrodynamic models by deriving the hydrodynamic equations from a
mean-field kinetic equation of filament dynamics.  In the
phenomenological approach the system is described in terms of a few
coarse-grained fields (conserved densities and broken symmetry
variables) whose dynamics is inferred from symmetry considerations.
The strength of this method is its generality.  Its drawback is that
for systems that are far from thermal equilibrium and therefore lack
constraints such as those provided by the fluctuation-dissipation
theorem or the Onsager relations, all the parameters in the equations
are undetermined. We have bridged the gap between microscopic models
and continuum theories by deriving the hydrodynamic equations through
a systematic coarse-graining of the microscopic dynamics. This
derivation provides an estimate of the various parameters in the
equations in terms of experimentally accessible quantities. We start
with a Smoluchowski equation for filaments in solution, where motor
proteins are described as active cross-linkers capable of exchanging
forces and torques between filaments. The active currents arising from
such motor-mediated exchange of forces and torques are obtained by
considering the kinematics of two filaments crosslinked by a single
active protein cluster that can rotate and translate at prescribed
rates as a rigid object relative to the filaments. The hydrodynamic
equations are then obtained by suitable coarse-graining of the
Smoluchowski equation.  This method yields a general form of the
hydrodynamic equations which incorporates all terms allowed by
symmetry, yet it provides a connection between the coarse-grained and
the microscopic dynamics.  In a series of earlier publications we have
described in details the derivation of the hydrodynamic equations for
filaments in a quiescent
solvent~\cite{tblmcm03,tblmcm04,tblmcm05,Ahmadi05,AAMCMTBL06}.  Here
we generalize this work by incorporating the flow of the solvent. This
is essential for describing the rheological properties of the
solution. A brief account of some of the results presented here have
been given elsewhere~\cite{TBLMCMPRL06}.
%%TBL

\section{Hydrodynamics of a solution of polar filaments}
\label{sec:m3} We consider a collection of rigid polar filaments in
a viscous solvent. The solution forms a quasi-two dimensional film, of
thickness much smaller than the length of the filaments.  Our goal is
to study the interplay of order and flow in controlling the phases and
the rheology of the system. The filaments diffuse in the solution and
can be crosslinked by both active and stationary protein clusters.
Active crosslinkers are small clusters of motor proteins that use
chemical fuel as an energy source to generate forces and torques on
the filaments, sliding and rotating filaments relative to each
other~\cite{howardBook,nedelec2}. In addition, other small proteins,
such as $\alpha$-actinins act as stationary crosslinkers and induce
filament alignment~\cite{Alberts_m}.

As in passive solutions of rigid filaments, the large scale
dynamics can be described in terms of a set of hydrodynamic
equations for continuum fields that relax on time scales much
longer than microscopic ones. These include the conserved variables
of the systems, as well as any field associated with broken
symmetries.  Various forms of these equations have been written
down phenomenologically by other authors. What distinguishes our
work from these phenomenological approaches is that we derive the
hydrodynamic equations from a mesoscopic model of coupled
motor-filament dynamics. This  allows us to estimate the various
parameters in the hydrodynamic equations (which are undetermined
in the phenomenological approach) and relate them to quantities
that can be controlled in experiments. To make contact with the
existing literature, we first present the equations and then
discuss their derivation via coarse-graining of a Smoluchowski
equation for rigid rods in a viscous solvent.

The conserved densities in a suspension of interacting filaments
(rods) in a solvent are the mass densities of filaments (rods)
$\rho_r(\bfr,t) = {\sf m} c(\bfr,t)$ and solvent $\rho_s(\bfr,t)$, and
the total momentum density ${\bf g}(\bfr,t)=\rho(\bfr,t)\bfv(\bfr,t)$
of the solution (rods+solvent), with $\bfv(\bfr,t)$ the flow velocity
and $\rho(\bfr,t)=\rho_s+\rho_r$ the total density. Here $c(\bfr,t)$
is the {\em number} density of rods and ${\sf m}$ the mass of a rod.
The conserved densities satisfy conservation laws, given by
\beqa
%&&\partial_t \rho_s=-\bm\nabla\cdot{\bf g}_s\;,\label{rhos_cons}\\
%&&\partial_t \rho_r=-\bm\nabla\cdot{\bf g}_r\;,\label{rhor_cons}\\
&&\partial_t \rho=-\bm\nabla\cdot{\bf g}\;,\label{rho_cons}\\
&&\partial_t c=-\bm\nabla\cdot{\bf J}\;,\label{c_cons}\\
&&\partial_tg_{i}+\partial_j(g_ig_j/\rho)=\partial_j\sigma_{ij}+
\rho F_i^{ext}\;, \label{mom_cons}
%0=\bm\nabla \cdot \left[ \right]
\eeqa
where ${\bf J}(\bfr,t)$ is the current density of rods
% $D_t = \partial_t + \rho^{-1} {\bf g} \cdot \nabla$
and ${\bf F}^{ext}$ the external force on the suspension.
%Addition of equations (\ref{rod_mom_cons}) and (\ref{fluid_mom_cons}) gives the momentum
%conservation equation. If the 2-fluid system is incompressible then
%$\rho+c=$constant and if we take the limit $\rho \gg c$ this
%approximates to $\bm\nabla \cdot {\bf v}_s=0$.
The stress tensor $\sigma_{ij}$ is the $i$-th component of the force
exerted by the surrounding fluid on a unit area perpendicular to the
$j$-th direction of a volume element of solution.  It includes all
forces on a volume of suspension exerted by the surrounding fluid. It
can be written as the sum of solvent and filament contributions as
\beq \sigma_{ij}=\sigma_{ij}^s+\sigma_{ij}^r\;. \eeq
The solvent contribution has the usual form appropriate for a
viscous fluid,
\beq
\sigma_{ij}^s=2\eta_0u_{ij}+(\eta_b-\eta_0)\delta_{ij}u_{kk}-\delta_{ij}\Pi_s(\rho)\;,
\eeq
where $\eta_0$ and $\eta_b$ are the shear and bulk viscosity of
the solvent, $\Pi_s(\rho)$ is the pressure of the solvent, and
$u_{ij}$ is the symmetrized rate of strain tensor,
\beqa u_{ij}=\frac{1}{2}\big(\partial_iv_j+\partial_jv_i\big)\;.
\eeqa
%
%\notes{Pressure implies incompressibility and therefore no bulk viscosity so we must make a choice here about what to use. I do not understand what you mean. A constant pressure implies incompressibility (in the absence of temperature fluctuations), but this is not what we have here, at least not yet. }
In the low Reynolds number limit we can ignore the inertial terms
on the left hand side of the solvent momentum equation,
Eq.~(\ref{mom_cons}).

In a solution of long filaments states with liquid crystalline order
are possible. Polar rods can form polarized and nematic states, both
characterized by orientational order, but with different symmetries
for the order parameters. Polar order in a fluid of $N$ rods is
characterized by a vector order parameter or polarization, ${\bf
  P}(\bfr,t)$, defined by
\beqa \label{polarization_def} {\bf
P}(\bfr,t)=\frac{1}{c(\bfr,t)}\Big\langle\sum_{\alpha=1}^N\uvec_\alpha(t)\delta(\bfr-\bfr_\alpha(t)\Big\rangle\;,
\eeqa
where $\bfr_\alpha$ is the position of the center of mass of the
$\alpha$-th rod and $\uvec_\alpha$ is a unit vector directed along the
polar direction. The angular bracket denote an ensemble average. The
polarization vector ${\bf P}$ can be written as
\beq {\bf P}(\bfr,t)=P(\bfr,t){\bf p}(\bfr,t)\;, \eeq
where the magnitude of the polarization $P(\bfr,t)$ is the scalar
order parameter and the unit vector ${\bf p}(\bfr,t)$ identifies the
direction of broken symmetry in the ordered state.

Nematic order is described by the conventional nematic order parameter
or alignment tensor, defined as
\beqa \label{Q_def}
Q_{ij}(\bfr,t)=\frac{1}{c(\bfr,t)}\Big\langle\sum_{\alpha=1}^N\Big(u_{\alpha
i}(t)u_{\alpha
j}(t)-\frac{1}{d}\delta_{ij}\Big)\delta(\bfr-\bfr_\alpha(t)\Big\rangle\;.
\eeqa
The subtracted part ensures that the order parameter vanishes in the
isotropic state in $d$ dimensions. The alignment tensor $Q_{ij}$ is
thus a traceless and symmetric second-order tensor field, with two
independent degrees of freedom in $d=2$. For uniaxial nematics the
alignment tensor takes the form
\beq \label{Q_def_macro}
Q_{ij}=S(\bfr,t)\Big[n_i(\bfr,t)n_j(\bfr,t)-\frac{1}{d}\delta_{ij}\Big]\;,
\eeq
where $S(\bfr,t)$ is the scalar nematic amplitude and ${\bf
  n}(\bfr,t)$ is the familiar nematic director. The nematic state has
orientational order ($S\not=0$) and it is invariant under inversion of
the director, i.e., for ${\bf n}\rightarrow -{\bf n}$. The polarized
state, in contrast, is not invariant for ${\bf p}\rightarrow -{\bf
  p}$. In a polarized state the alignment tensor $Q_{ij}$ is slaved to
the polarization and acquires a nonzero value, with ${\bf n}={\bf p}$.

The dynamical equations for polarization and alignment tensor have the
form (for simplicity we give the form for $d=2$ only)
\beqa &&\label{Pi}
D_t (cP_i)=cF_i(\bm\kappa,{\bf P})-\partial_jJ_{ij}-R_i\;,\\
&&
\label{Qij}
D_t (cQ_{ij})=cF_{ij}(\bm\kappa,{\bf Q})
-\partial_kJ_{ijk}-R_{ij}\;, \eeqa
where $D_t = \partial_t + \bfv \cdot \bnabla $, $\bm\kappa$ is the
rate of strain tensor, $\kappa_{ij}=\partial_jv_i$, and
%
%\begin{widetext}
%
\beqa \label{Fi} F_i(\bm\kappa,{\bf
P})&=&-\omega_{ij}P_j+\lambda_P u_{ij} P_j-\frac{5}{4} u_{kk}
P_i \quad , \quad \mbox{or} \nonumber\\
\bfF(\bm\kappa,{\bf
P})&=&\frac{1}{2}(\bnabla\times\bfv)\times {\bf
P}+\lambda_P\big[\bnabla\bfv +(\bnabla\bfv)^T\big]\cdot{\bf
P}-\frac{5}{4}(\bnabla\cdot\bfv){\bf P}\;,\eeqa
%
%\notes{mixture of general and specific $d$ in this formula}
\beqa \label{Fij}F_{ij}(\bm\kappa,{\bf Q})&=&
-\big(\omega_{ik}Q_{kj}+\omega_{jk}Q_{ki}\big)
+\frac{1}{3}\Big(u_{ik}Q_{kj}+u_{jk}Q_{ki}-\delta_{ij}u_{kl}Q_{kl}\Big)\nonumber\\
&&-\frac{4}{3}u_{kk}Q_{ij}+\lambda
\Big(u_{ij}-\frac{1}{2}\delta_{ij}u_{kk}\Big)\;.\eeqa
%\end{widetext}
%
Here $\lambda_P$ and $\lambda$ are the flow alignment parameters in
the polarized and nematic states, respectively, and
\beqa
\omega_{ij}=\frac{1}{2}\big(\partial_iv_j-\partial_jv_i\big)\;.
\eeqa
The low density derivation based on the Smoluchowski equation
described below gives $\lambda_P=1/2$ and $\lambda=1/(2S_0)$, with
$S_0$ the magnitude of the nematic order parameter.  Since typically
in the nematic state even quite far from the I-N transition, $S_0 \ll
1$, we expect $\lambda > 1$, as required for flow alignment. In
addition it is well known that deep in the nematic phase, higher order
correlations can further increase the value of $\lambda$.  In the
following we will treat both $\lambda_P$ and $\lambda$ as unknown
parameters. The first terms on the right hand side of Eqs.~(\ref{Pi})
and (\ref{Qij}) generalize the convective derivative on the left hand
side of the equation to the case of long, thin rods. These are
standard terms in nematohydrodynamics that have been derived from a
microscopic model before~\cite{Kuzuu1983}.  Different values for some
of the numerical coefficients are reported in the literature,
depending on the closure scheme used in evaluating various angular
averages.  The second and third terms on the right hand side of
Eqs.~(\ref{Pi}) and (\ref{Qij}) represent translational and rotational
currents, including contributions from diffusion, excluded volume, and
both stationary and active cross-linkers. The relaxation of the order
parameters is controlled by the rotational currents and is
non-hydrodynamic. In contrast, the relaxation of the broken symmetry
variables ${\bf p}$ and ${\bf n}$ is controlled by hydrodynamic
Goldstone modes, as appropriate in ordered states.

The long wavelength description of the solution is then given by the
five equations (\ref{rho_cons}-\ref{mom_cons}) and
(\ref{Pi}-\ref{Qij}). To close the hydrodynamic equations we must
derive the constitutive equations for the fluxes (${\bf J}$, $J_{ij}$,
$J_{ijk}$), the rotational currents ($R_i$ and $R_{ij}$), and the
filament contribution to the stress tensor, $\sigma_{ij}^r$, as
functions of the system's properties (density, filament concentration
and order parameters) and of the driving forces (applied mechanical
stresses and activity, as measured by the ATP consumption rate). This
derivation is carried out below by adapting methods from polymer
physics appropriate for a dilute solution of rigid rods. Although the
specific expressions obtained by this method for the parameters in the
hydrodynamic equations only apply at low concentration of filaments,
the structure of the equations is general and remains the same at high
density.

\section{Derivation of hydrodynamic constitutive equations}
\label{sec:m4}

Our goal is to derive the constitutive equations for the various
hydrodynamic currents and stresses starting from a semi-microscopic
model of the dynamics of single filaments coupled pairwise by active
and stationary crosslinkers. The filaments are modeled as rigid rods
of fixed length $ l $ and diameter $b \ll l $ immersed in a viscous
solvent. They diffuse independently in the solvent and interact via
excluded volume. In addition, filaments can be coupled pairwise by
both stationary and active crosslinkers that generate additional
active currents. Active crosslinkers are described as rigid links that
can walk along the filaments towards the polar end at a prescribed
rate $u(s)$ proportional to the rate of ATP consumption. Generally
$u(s)$ varies with the point $s$ of attachment along the filament $(0
\le s \le l)$.  Both active and stationary cross-links also mediate
the exchange of torques between the filaments by acting as torsional
springs of prescribed stiffness, $\kappa$. Our goal is to obtain a
coarse-grained description of the system where all the parameters in
the hydrodynamic equations are characterized in terms of $u(s)$,
$\kappa$, and the density of crosslinkers. Collective effects arising
from multiple crosslinkers are neglected and the density of
crosslinkers is assumed constant for simplicity.  We also neglect the
dynamics of crosslinkers binding and unbinding which occurs on faster
time scales than those of interest here, so that we can treat a
constant fraction of them as bound. The dynamics of crosslinkers
binding and unbinding was considered for instance in
Ref.~\cite{Sankararaman04} and it was found that varying the rates of
motor binding and unbinding did not affect the nonequilibrium steady
states of the active solution. The derivation of the active
contributions to the various fluxes has been presented
elsewhere~\cite{AAMCMTBL06} and will be summarized here for
completeness. We also present novel results on the evaluation of the
filament contribution to the stress tensor up to terms of first order
in gradients of the hydrodynamic fields.

To proceed, we also make a series of simplifying assumptions on the
dynamics of the solution. First, we assume that the friction between
filaments and solvent is large and the filaments move at the flow
velocity ${\bf v}={\bf g}/\rho$ of the solution. In many fluid
mixtures internal friction mechanisms are so strong that the flow
velocities of the two components relax on microscopic time scales to
the common value ${\bf v}$. There are situations, however, where the
relaxation time of the relative momenta of the two species is slow
enough to have a significant influence even on hydrodynamic time
scales. In this case a two-fluid description is appropriate and
useful. Such a "two-fluid model" of the system (rods and fluid
background) will be described elsewhere, where we will show under
which conditions one approaches the one-fluid model (which is always
the true hydrodynamic limit).

We also limit ourselves to the case of incompressible solutions, with
$\rho=\rho_s+\rho_r={\rm constant}$, which requires
\beq \label{divv}\bm\nabla\cdot\bfv=0\;. \eeq
Finally, we neglect fluid inertial effect compared to the
frictional forces between the colloidal rods and the solvent. In
this limit the momentum equation (\ref{mom_cons}) reduces to the
Stokes equation
\beq\label{Stokes}\partial_j\sigma_{ij}=-\rho F^{\rm ext}_i\;,
\eeq
or, in the absence of external forces,
\beq \label{Stokes_expl} \eta_0\nabla^2v_i
-\partial_i\Pi_s=-\partial_j\sigma_{ij}^r\;. \eeq
Equation (\ref{Stokes_expl}) shows that the flow velocity of the
solution is determined by the stress introduced by the filaments.  In
turn, the forces that the filaments exert on each other and on the
solvent depend on the flow of the suspension in which they are
immersed and the problem must be solved self-consistently.

\begin{figure}
\center \resizebox{0.49\textwidth}{!}{
  \includegraphics{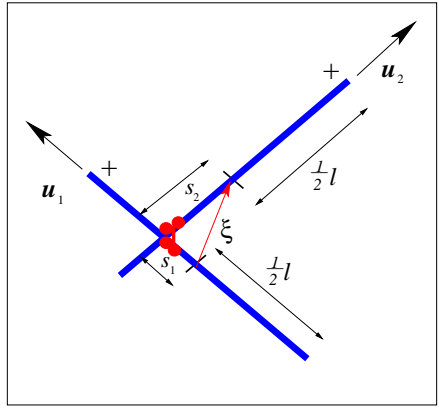}}

\caption{The geometry of overlap between two interacting filaments
of length $l$ cross-linked by an active cluster. The cross-link is
a distance $s_1,(s_2)$ from the center of mass of filament $1(2)$.
The distance between centers is $\bm{\xi}= {\bf r}_2-{\bf r}_1
= s_1 \nvec_1 - s_2 \nvec_2$.}
\label{Fig:2filam_coord}       % Give a unique label
\end{figure}

The dynamics of a dilute suspension of rods in the presence of a
macroscopic flow field $\bfv(\bfr)$ can de described by the
Smoluchowski equation for the probability distribution
$\chat(\bfr,\uvec,t)$ of rods with center of mass at $\bfr$ and
orientation $\uvec$ at time $t$. The Smoluchowski equation describes
the mean-field Brownian dynamics of extended colloidal particles at
low Reynolds number, under the assumption that the particles
velocities have equilibrated on microscopic time scales to a local
Maxwell-Boltzmann distribution at a temperature
$T_a$~\cite{langer,lma01}.  The effective temperature $T_a$
incorporates nonthermal noise sources as may arise from fluctuations
in motor concentration and ATP consumption rate. The Smoluchowski
equation is given by
\begin{equation}
\partial_t \chat
%+ \bnabla \cdot ({\bf v} \chat) + {\cal R} \cdot \left( \bomega \chat \right)
+ \bnabla \cdot {\bf
{J}}_c + {\cal {R}} \cdot {\cal {J}}_c =0\;, \label{smoluchowski}
\end{equation}
where ${\cal {\bf R}}=\uvec\times\partial_{\uvec}$ is the rotation
operator. The {\em translational} probability current, ${\bf {J}}_c$,
and the {\em rotational} probability current, $\bm{\mathcal {{J}}}_c$,
are given by
\begin{eqnarray}
\label{current}
{J}_{ci}&=&\chat v_i -D_{ij}\nabla_{j}\chat-\frac{D_{ij}}{k_BT_a}\chat~\nabla_{j}U_{\rm
ex}+{J}_{ci}^{\rm A}\;, \\
\label{current2}{\mathcal {{J}}}_{ci} &=& \chat \omega_i -D_r {\cal R}_i \chat -
{D_r \over k_B T_a} \chat {\mathcal{R}}_iU_{\rm
ex}+\mathcal{{J}}_{ci}^{\rm A}\;,
\end{eqnarray}
where $\omega_i = \epsilon_{ijk}\uhat_j \uhat _l \partial_l v_k$.
Also
$D_{ij}=D_\parallel\uhat_i\uhat_j+D_\perp\big(\delta_{ij}-\uhat_i\uhat_j\big)$
is the translational diffusion tensor and $D_r$ is the rotational
diffusion rate. For a low-density solution of long, thin rods
$D_\perp=D_\parallel/2\equiv D/2$, where $D=k_BT_a\ln( l
/b)/(2\pi\eta_0 l )$, and $D_r=6D/l^2$. The potential $U_{\rm ex}$
incorporates excluded volume effects which give rise to the nematic
transition in a solution of hard rods. It can be written by
generalizing the Onsager interaction to inhomogeneous systems as
$k_BT_a$ times the probability of finding another rod within the
interaction area of a given rod. In two dimensions this gives
\begin{eqnarray}
\label{Vex} U_{\rm ex}(\bfr_1,\uvec_1)= k_BT_a\int
d\uvec_2\int_{s_1 s_2}|\uvec_1\times\uvec_2|~\chat({\bf
r}_1+\bm{\xi},\uvec_2,t)\;,
\end{eqnarray}
where $s_i$, with $-l/2\leq s_i\leq l/2$, parameterizes the position
along the length of the $i$-th filament, for $i=1,2$, and
$\int_{s_i}...\equiv \int_{-l/2}^{l/2}~ds_i...\equiv\langle
..\rangle_{s_i}$. The filaments are constrained to be within each
other's interaction volume, i.e., in the thin rod limit $b \ll l$
considered here, have a point of contact. The factor
$|\uvec_1\times\uvec_2|$ represents the excluded area of two thin
filaments of orientation $\uvec_1$ and $\uvec_2$ touching at one
point~\cite{DoiEdwards}. Finally, $\bm{\xi}={\bf r}_2-{\bf r}_1 \simeq
\uvec_1s_1-\uvec_2s_2$, is the separation of the centers of mass of
the two rods. The translational and rotational active current of
filaments with center of mass at ${\bf r}_1$ and orientation along
$\uvec_1$ are written as

\begin{eqnarray}
&&{\bf {J}}_c^{\rm A}({\bf r}_1,\uvec_1)=\chat({\bf
r}_1,\uvec_1,t)b^2m\int_{\uvec_2}\int_{s_1
s_2}\bfv_a(1;2)  \chat({\bf
r}_1+\bm\xi,\uvec_2,t)\;,\label{Jtact}\\
&&\bm{\mathcal J}_c^{\rm A}({\bf r}_1,\uvec_1)=\chat({\bf
r}_1,\uvec_1,t)b^2m\int _{\uvec_2}\int_{s_1
s_2} \bm\omega_a(1;2) \chat({\bf
r}_1+\bm\xi,\uvec_2,t)\;, \;,\label{Jract}
\end{eqnarray}
where $m$ is the density of bound crosslinkers and
$(1;2)=(s_1,\uvec_1;s_2,\uvec_2)$. Finally, ${\bf v}_a(1;2)$ and
$\bm\omega_a(1;2)$ are the translational and rotational velocities,
respectively, that filament 1 acquires due to the crosslinker-mediated
interaction with filament 2, when the centers of mass of the two
filaments are separated by $\bm\xi$ (see Fig.~\ref{Fig:2filam_coord}).

\begin{figure}
\centering
\includegraphics[height=4cm]{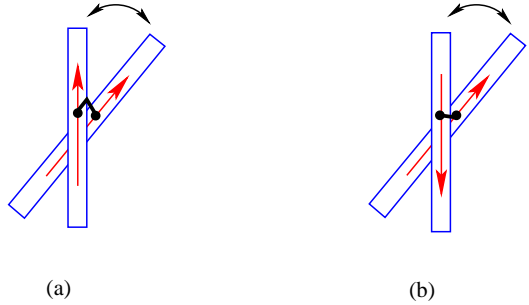}
%\picplace{5cm}{2cm} % Give the correct figure height and width in cm
%
\caption{Polar and nonpolar clusters interacting with polar
  filaments. Assuming that clusters always bind to the smallest angle,
  polar clusters  bind only to filaments in
  configuration (a) while non-polar clusters bind to both
  configurations equally.}
\label{Fig:PolarNonpolar}       % Give a unique label
\end{figure}

The derivation of the form of the active velocities in terms of motor
parameters (the stepping rate $u(s)$ and the torsional stiffness
$\kappa$) has been discussed in detail
elsewhere~\cite{tblmcm05,AAMCMTBL06}. The angular velocity is
\beq \label{omegaa}{\bm\omega}_a = 2\left[\gamma_P +
\gamma_{NP}\left(\uvec_1 \cdot \uvec_2 \right)  \right] \left(
\uvec_1 \times \uvec_2 \right)\;, \eeq
with $\gamma_P$ and $\gamma_{NP}$ motor-induced rotation rates due to
polar and nonpolar crosslinkers, respectively (see
Fig.~\ref{Fig:PolarNonpolar}). The motor-induced translational
velocity has the form $\bfv_a(1;2)={1\over 2} \bfv_{r}+{\bf V}_{m}$,
with~\cite{AAMCMTBL06}

\begin{eqnarray}
\label{vel_eq2}
%{\bf v}_{r}&=&\left(\beta-\alpha{\textstyle{\frac{s_1+s_2}
%{2}}}\right)\left(\uvec_2-\uvec_1\right) + \alpha
%{\textstyle\frac{s_1-s_2}{2}}(\uvec_2+\uvec_1)\;, \nonumber\\
{\bf v}_{r}&=&\frac{\tilde\beta}{2}(\uvec_2-\uvec_1)
+\frac{\tilde\alpha}{2l}\bm\xi\;, \nonumber\\
\label{V} {\bf V}_{m} & =& A \left(\uvec_2 + \uvec_1 \right) + B
\left(\uvec_2 - \uvec_1 \right)\;,\nonumber
\end{eqnarray}
where $\tilde\alpha=\alpha(1+\uvec_1\cdot\uvec_2)$ and
$\tilde\beta=\beta(1+\uvec_1\cdot\uvec_2)$.  The expressions for $A$
and $B$ have been obtained in ~\cite{AAMCMTBL06} using momentum
conservation. For long thin rods with
$\zeta_\perp=2\zeta_\parallel\equiv 2\zeta$, to leading order in
$\uvec_1 \cdot \uvec_2$, we find $A=-[\beta-\alpha(s_1+s_2)/2]/12$ and
$B=\alpha(s_1-s_2)/24$.

The rotational rates, $\gamma_P$ and $\gamma_{NP}$, and the active
velocities $\alpha$ and $\beta$ can be related to the torsional
stiffness $\kappa$ of the crosslinkers and to the rate $u(s)$ at which
a motor cluster attached at position $s$ steps along a filament
towards the polar end. This rate will in general depend on the point
of attachment $s$, due for instance to crowding or stalling of motors
near the polar end. The mean (averaged along the filament) stepping
rate $u_0=\langle u(s)\rangle/ l $ is simply proportional to the mean
rate $R_{ATP}$ of ATP consumption via a the characteristic step
length, which we take of order of the thickness $b$ of the filaments,
$u_0\sim bR_{ATP}$.
%%TBL
We emphasize, however, that in general the stepping rate $u(s)$ (and
other active parameters) may be a {\em non-linear} and possibly even
non-monotonic function of the rate of ATP consumption, $R_{ATP}$.
%%TBL

In our model there are three coupled mechanisms for
crosslinker-induced filament dynamics, described by the parameters
$\alpha$, $\beta$ and the rotational rates, $\gamma_P$ and
$\gamma_{NP}$.  The first is the bundling of filament of similar
polarity at a rate $\alpha$ given by~\cite{tblmcm05}
\beq\label{alpha_par} \alpha=
\int_{- l /2}^{ l /2}\frac{ds}{ l }~\frac{s}{ l }~u(s)\approx
u_0(b/ l )\;, \eeq
where the last approximate equality applies in situations where $u(s)$
exhibits strong spatial variations on length scales of order $b$, as
may arise for instance from motor stalling at the polar
end~\cite{tblmcm05}. It is apparent from Eq.~(\ref{alpha_par}) that
$\alpha=0$ if $u(s)$ is constant.  Bundling is driven by the
contractile nature of motor clusters and in our mean field model
requires spatial inhomogeneities in the rate at which motors step
along the filaments. As we will see below, it tends to build up
density inhomogeneities and is the main pattern-forming mechanism. The
second mechanism of motor-induced dynamics will be referred to as
"polarization sorting", although in general it involves coupled
filament rotation and spatial separation of filaments of different
polarity. It occurs at the rate
\beq\label{beta_par} \beta=
\int_{- l /2}^{ l /2}\frac{ds}{ l }u(s)=u_0\;, \eeq
and vanishes for aligned filaments. This mechanisms is especially
important in the polar state where it allows for terms in the
hydrodynamic equations corresponding to convection of filaments along
the direction of mean polarization and it provides the mechanism for
the transition to a state with spontaneous flow~\cite{Voituriez05}.
Finally, motor-induced filament rotations occur at rates $\gamma_P$
and $\gamma_{NP}$ for crosslinkers that preferentially bind to
filaments of the same orientation ($\gamma_P$) or regardless of their
orientation ($\gamma_{NP}$).  As discussed in Ref.~\cite{AAMCMTBL06},
we estimate
\beq \label{gamma_par}
\gamma_P\sim\gamma_{NP}\sim\frac{\kappa}{\zeta_r}\;, \eeq
with $\zeta_r=k_BT_a/D_r$ a rotational friction. In general both
active and stationary crosslinkers may induce rotation and be either
polar or apolar in nature. In the following we will restrict ourselves
for simplicity to the case where all polar cross-linkers are active
motor clusters (of density $m_a$), while all apolar crosslinkers are
stationary (of density $m_s$). In practice we do expect this to be
often the case.  The rotational rate $\gamma_P$ will then depend on
ATP consumption, with $\gamma_P\sim R_{ATP}$, while we expect
$\gamma_{NP}$ to be essentially independent of it.
%%TBL
As mentioned above, the various active parameters may be non-linear
and even non-monotonic functions of $R_{ATP}$. However, these effects
will not be considered here.
%%TBL
%
%\begin{eqnarray}
%A &=& - \left({1 \over 12} \right) \left({\textstyle{\frac{1 - \uvec_1 \cdot \uvec_2}
%{1 - \uvec_1 \cdot \uvec_2/3 }}}\right) \left[\beta-\alpha{\textstyle{\frac{s_1+s_2}{2}}}\right] \\
%B &=&  \left({1\over 12} \right) \left( {\textstyle{1 + \uvec_1
%\cdot \uvec_2 \over 1 + \uvec_1 \cdot \uvec_2/3
%}}\right)\left[\alpha{\textstyle{\frac{s_1-s_2}{2}}}\right]\;,
%\end{eqnarray}
%

From the Smoluchowski equation, Eq.~(\ref{smoluchowski}), we obtain the
hydrodynamic equations for filament concentration, polarization and
alignment tensor by truncating the exact moment expansion of
$\chat(\bfr,\uvec,t)$ as
\begin{equation}
\chat(\bfr,\uvec,t)=\frac{c(\bfr,t)}{2 \pi}\Big\{1+2 {\bf
P}(\bfr,t)\cdot\uvec+4Q_{ij}(\bfr,t)\hat{Q}_{ij}(\uvec)+\ldots \Big\}\;,
\label{tens_expand}\end{equation}
with $\hat{Q}_{ij}(\uvec)=\uhat_i\uhat_j-\frac{1}{2}\delta_{ij}$
and keeping only the first three moments,
\begin{eqnarray}
&&\int d\uvec ~\chat(\bfr,\uvec,t)=c(\bfr,t)\;\; \mbox{(density)}, \nonumber\\
&&\int d\uvec ~\uvec~\chat(\bfr,\uvec,t)=c(\bfr,t){\bf P}(\bfr,t)\;\;\mbox{(polarization)}, \\
&&\int d\uvec
~\hat{Q}_{ij}(\uvec)~\chat(\bfr,\uvec,t)=c(\bfr,t)Q_{ij}(\bfr,t)\;\;\mbox{(nematic
order)} \, . \nonumber \label{orient_moments}\end{eqnarray}
The details of the calculation, which involves using a small gradient
expansion for the filament probability distribution and evaluating
angular averages, are given in \cite{AAMCMTBL06}, where the full
expression for the various fluxes and rotational currents are also
displayed. The resulting hydrodynamic equations in the isotropic and
ordered phases will be given below.

\section{Stress tensor of an active solution}
\label{sec:m5}

In this section we derive the constitutive equation for the filament
contribution to the stress tensor of an active suspension of polar
filaments.  An important difference as compared to passive solutions
is that in active systems stresses can be induced not just by
externally applied mechanical deformations (yielding
$\kappa_{ij}\not=0$), but also by motor activity which maintains the
system out of equilibrium by supplying energy at a rate $R_{ATP}$.

In the limit where inertial effects may be ignored (low Reynolds
number) and in the absence of external forces, momentum conservation
is described by Eq.~(\ref{Stokes_expl}), with $\bnabla \cdot \bfv =0$.
Using standard methods from polymer physics, the filament contribution
to the divergence of the stress tensor of a dilute suspension of hard,
thin rods can be written as
\begin{eqnarray}
\label{fils_stress_micro}
 \bnabla \cdot \bsigma^r & =& -\int_{\bxi}
\int_{\uhat} \chat({\bf r}-\bxi,\uvec,t) \Big\langle\delta \left(
{\bxi}-s \uvec \right) {\bf {\cal F}}^h(s)\Big\rangle_s \; \; ,
\end{eqnarray}
where ${{\bf {\cal F}}}^h(s)$ is the hydrodynamic force per unit
length exerted by the suspension on a rod at position $s$ along the
rod. It arises from interactions with other filaments and proteins and
with solvent molecules. It depends implicitly upon direct interactions
between the rods, as well as on hydrodynamic interactions mediated by
the solvent.

The hydrodynamic force density on a rigid rod suspended in a viscous
solvent can be expressed in terms of the force and torque at its
center of mass.  A sketch of the derivation is given in Appendix
\ref{sec:mappA}. Further details of similar calculations can be found
%%TBL
in \cite{DoiEdwards,dhont03}.
%%TBL
We find that the stress due to the filaments can be written in the
form (to ${\cal O}(\nabla^2)$)
\begin{eqnarray}
\nabla \cdot {\bm \sigma}^r (\bfr,t) &=& \int_{\uvec} \chat({\bf
r},\uvec,t) \bfF^h({\bf r},\uvec,t)  \nonumber \\ && -
\int_{\uvec} \big\langle \big(\frac{s}{l}\big)^2\left(\frac{\uvec
\cdot \nabla}{l} \right)\chat({\bf r},\uvec,t){\bm \tau}^h ({\bf
r},\uvec,t)\big\rangle_s\;. \label{eq:stress_ftau}\end{eqnarray}
In the absence of inertial effects, the \emph{total} hydrodynamic
force, ${\bf F}^h(\bfr,\uvec,t)$, exerted by the suspension \emph{on
  the center of mass} of a rod can be found from the condition that
all forces acting on the rod must balance.  The solvent flow field on
a given segment of a rod is calculated using a decoupling
approximation where the hydrodynamic coupling to other segments of the
same rod are treated explicitly within the Oseen approximation, while
the hydrodynamic effects of other rods enter in the determination of a
self-consistent value for the flow velocity of the solvent, yielding,
\beq {\bf F}^h(\bfr,\uvec,t)=k_BT_a\bnabla \ln \chat+\bnabla
U_{\rm ex}-{\bf F}_a\;, \eeq
where $-k_BT_a\bnabla \ln \chat$ is the Brownian force, $-\bnabla
U_{\rm ex}$ is the force due to the direct interaction of the rod with
other rods (in this case, via excluded volume) and ${\bf F}_a$ is the
active force that can be written as
\beq F_{ai}=\zeta_{ij}(\uvec)J_{ci}^A/\hat{c}\;. \eeq
The rod friction tensor $\zeta_{ij}(\uvec)$ is proportional to the
inverse of the rod diffusion tensor $D_{ij}(\uvec)$, with
\beqa \zeta_{ij}(\uvec)&=&k_B T_a \left[{\bf
D}^{-1}(\uvec)\right]_{ij}\nonumber\\
&&=\zeta_\parallel
\uhat_i\uhat_j+\zeta_\perp(\delta_{ij}-\uhat_i\uhat_j)\;, \eeqa
with $\zeta_\parallel=2 \pi \eta_0 l / \ln(l/b)$ and
$\zeta_\perp=2 \zeta_{\|}$.
%
% In order to evaluate the right
%hand side of Eq.~(\ref{fils_stress_micro}) in terms of the
%filament concentration and the orientational order parameters, we
%need, however, an expression for the hydrodynamic force
%distribution along a rod's entire axis, rather than just the force
%at its the center of mass. The derivation of such a force is
%sketched in Appendix . Further details can be found for instance
%in .
Similarly the total hydrodynamic torque is given by
\begin{eqnarray}
\label{torque} {\bm \tau}^h ({\bf r},\uvec,t) &=& \left[ k_B T_a
{\cal R} \ln c + {\cal R} U_x
- {\bm\tau}_{a}\right] \times \uvec \nonumber \\
&& - {\zeta_\perp \over 2} \uvec\uvec (\uvec \cdot \nabla) \cdot
\bfv(\bfr)\;,
\end{eqnarray}
with $\bm\tau_a = \zeta_r {\cal J}_c^A/ \hat{c}$ the active torque.
The last term on the right hand side of Eq.~(\ref{torque}) is a
viscous contribution to the stress proportional to the velocity
gradient.

The rod contribution to the stress tensor can now be evaluated
explicitly using the truncated moment expansion for $\chat(\bfr,\uvec,
t)$ given in Eq.~(\ref{tens_expand}). When evaluating the active
contributions to the stress tensor, only terms up to first order in
$\uvec_1\cdot\uvec_2$ are retained in the active force
$\bm\zeta(\uvec_1)\cdot{\bf v}_a(1;2)$ exerted by a motor cluster on
the filament. This approximation only affects the numerical values of
the coefficients in the stress tensor, not its general form.

For simplicity, we consider solutions in the presence of a constant
velocity gradient, $\kappa_{ij}$, and with a uniform mean rate of ATP
consumption. We allow for spatial inhomogeneities in the filament
concentration and orientational order parameters and evaluate the
stress tensor up to first order in gradients of these hydrodynamic
fields. The deviatoric part
$\tilde{\sigma}_{ij}=\sigma_{ij}-(1/2)\delta_{ij}\sigma_{kk}$ of the
stress tensor of the filaments is
%(1-active forces, 2-passive torques and 3-passive forces)
%
\begin{equation}
\tilde{\sigma}_{ij}^r (\bfr,t) = \tilde{\sigma}_{ij}^A (\bfr,t) +
\tilde{\sigma}_{ij}^v (\bfr,t)\; ,
\end{equation}
with
\begin{eqnarray}
\tilde{\sigma}_{ij}^A & =& 2 k_B T_a c\Big(1 - {c \over c_{IN}}
\Big)  Q_{ij}
-k_B T_a { c^2 \over c_{IP}} \Big( P_i P_j - {1 \over 2}  \delta_{ij}P^2 \Big)\nonumber \\
&&+ m_ab^2\alpha \frac{k_B T_a l^3}{72D} c^2\Big({4  \over 3}
Q_{ij} + P_i
P_j - {1 \over 2}\delta_{ij}   P^2 \Big)\nonumber\\
&&+m_ab^2\beta\frac{k_B T_a l^4}{216D}c^2\Big[\partial_j
P_i-\frac{1}{2}\delta_{ij}\bnabla\cdot{\bf
P}-\frac{1}{4}\Big(\partial_i P_j-\partial_j P_i\Big)\Big]\;,
\label{act_stress}
\end{eqnarray}
where $c_{IP}=D_r/(m_a b^2\gamma_Pl^2)$, and $c_{IN}=c_N/[1+c_N l^2m_s
b^2\gamma_{NP}/(4D_r)]$ are the densities for the isotropic-polarized
(IP) and isotropic-nematic (IN) transition, respectively, at finite
density of active polar motor clusters ($m_a$) and stationary nonpolar
crosslinkers ($m_s$)\cite{AAMCMTBL06}. Finally, $c_N=3\pi/(2 l^2)$ is
the density of the IN transition in passive systems. There are three
types of contributions to the active part of the stress tensor.  The
first consists of the first two terms on the right hand side of
Eq.~(\ref{act_stress}). These are equilibrium-like terms, in the sense
that they have the same structure one would obtain in a nematic and
polar passive fluid, respectively, with the transition densities
replaced by their active values. In particular, the first term on the
right hand side of Eq.~(\ref{act_stress}) should be compared to the
corresponding contribution for
%%TBL
isotropic ($c < c_N$) passive solutions,
%%TBL
$\tilde{\sigma}_{ij}^P = 2 k_B T c\Big(1 - {c \over c_{N}} \Big)
Q_{ij}$.  The third term is a homogeneous nonequilibrium contribution
that remains nonzero even for $\kappa_{ij}=0$. This "spontaneous
stress" arises from activity and is proportional to the ATP
consumption rate that acts as an additional driving force and can
build up stresses even in the absence of mechanical deformations. This
term is generated by motor-induced filament bundling and it is
proportional to the bundling rate, $\alpha$. It would therefore vanish
in the absence of spatial inhomogeneities in the motor stepping rate.
Finally, the fourth term contains active contributions proportional to
gradients of the polarization (we have omitted here terms of linear
order in the gradients containing both polarization and alignment
tensor. The full expression for the stress tensor can be found in
Appendix A). These stresses are generated by motor-induced filament
sorting and are proportional to $\beta$.  They are important only in
the polarized phase, where we expect they will play an important role
in enhancing the relaxation of longitudinal fluctuations of the
filaments and the corresponding relaxation of shear via reptation.

Finally, the viscous contribution to the stress is
\begin{eqnarray}
\tilde{\sigma}_{ij}^v &=& {l c \zeta_\perp \over 48} \Big[{1 \over
2} \big(u_{ij} -{\textstyle{1\over 2}}\kappa_{kk}\delta_{ij}\big)
+ {1 \over 3} \big( Q_{ij} \kappa_{kk}-\delta_{ij}
\kappa_{kq} Q_{qk} \big) \nonumber \\
&& + \frac{2}{3}\Big( u_{ik}Q_{kj} + u_{jk}Q_{ki}\Big)
\Big]\;,\label{visc_stress}
\end{eqnarray}

\section{Homogeneous states of a quiescent solution}
\label{sec:m6}

We first examine the case of a quiescent suspension, with $\bfv=0$. We
consider a system with a concentration $m_a$ of active, polar motor
clusters and a concentration $m_s$ of stationary nonpolar
crosslinkers. For convenience we define a dimensionless parameter
$\mu_a$ measuring activity as
\beq \label{muA}\mu_a=m_ab^2\frac{\gamma_P}{D_r}\sim m_a
R_{ATP}\;, \eeq
where $D_r$ is the rods' rotational diffusion constant and we have
assumed that $\gamma_P\sim R_{ATP}$. We also introduce a dimensionless
parameter $\mu_s$ measuring the effect of stationary crosslinkers as
\beq \label{mus} \mu_s=m_sb^2\frac{\gamma_{NP}}{D_r}\;, \eeq
and assume that $\mu_s$ is essentially independent of the ATP
consumption rate.
%%TBL
The bulk states of the system are determined by the solution of the
homogeneous hydrodynamic equations containing only those terms that
are of zeroth order in the gradients. This is the most coarse-grained
description of the system.  More detailed descriptions that
incorporate slowly varying spatial variations can then be developed by
including gradient terms in the hydrodynamics.
%%TBL
The possible homogeneous states of the system
are obtained as the stationary solution of the homogeneous
hydrodynamic equations for filament concentration, polarization
and alignment tensor, setting all gradient terms equal to zero. In
this case the filament concentration is constant, $c=c_0$, and
only contributions from rotational currents survive in  equation
for the polarization and the alignment tensor, which are given by
\begin{eqnarray}
\label{st_P_dim}
\partial_tP_i=-D_r\big[1-\mu_a c_0\big]P_i+
D_r\Big[4c_0/c_N+(\mu_s-2\mu_a)c_0\Big] Q_{ij}P_j\;,
\end{eqnarray}
\begin{eqnarray}
\label{st_Q_dim}\partial_tQ_{ij}=-D_r\Big[4(1- c_0/ c_N)-\mu_s
c_0\Big]Q_{ij} +2D_r\mu_a c_0
\Big(P_iP_j-\frac{1}{2}\delta_{ij}P^2\Big)\;,
\end{eqnarray}
where all filament densities are measured in units of $l^2$, and $
c_N=3\pi/2$.

There are three possible homogeneous stationary states for the system,
obtained by solving Eqs.~(\ref{st_P_dim}) and (\ref{st_Q_dim}) with
$\partial_tP_i=0$ and $\partial_t Q_{ij}=0$.  These are:
\begin{eqnarray}
&&{\rm isotropic}\hspace{0.1in}{\rm state}\hspace{0.1in}{\rm
(I):}\hspace{0.3in}
P_i=0\hspace{0.2in}Q_{ij}=0\;,\nonumber\\
&&{\rm nematic}\hspace{0.1in}{\rm state}\hspace{0.1in}{\rm
(N):}\hspace{0.3in}
P_i=0\hspace{0.2in}Q_{ij}\not=0\;,\nonumber\\
&&{\rm polarized}\hspace{0.1in}{\rm state}\hspace{0.1in}{\rm
(P):}\hspace{0.3in}
P_i\not=0\hspace{0.2in}Q_{ij}\not=0\;.\nonumber
\end{eqnarray}
At low density the only solution is $P_i=0$ and $Q_{ij}=0$ and the
system is isotropic (I). The homogeneous isotropic state can become
unstable at high filament and/or motor density, as described below.

To discuss the instabilities it is convenient to measure time in units
of $D_r^{-1}$ and rewrite Eqs.~(\ref{st_P_dim}) and (\ref{st_Q_dim})
in a more compact form as
\begin{eqnarray}
\label{st_P}&& \partial_tP_i=-a_1P_i+b_1 c_0 Q_{ij}P_j\;,
\end{eqnarray}
\begin{eqnarray}
\label{st_Q}&&\partial_tQ_{ij}=-a_2Q_{ij}+b_2 c_0
\Big(P_iP_j-\frac{1}{2}\delta_{ij}P^2\Big)\;.
\end{eqnarray}
The coefficients $a_1$, $b_1$, $a_2$, and $b_2$ are given by
\begin{eqnarray}
&& a_1=1-\mu_ac_0\;,\\
&&a_2=4\big[1- c_0/ c_{N}-\mu_sc_0/4\big]\;,\\
&&b_1=4/c_N+\mu_s-2\mu_a\;,\\
&&b_2=2\mu_a\;.
\end{eqnarray}

In the absence of crosslinkers ($\mu_s=\mu_a=0$), no homogeneous
polarized state with a nonzero mean value of ${\bf P}$ is obtained.
There is, however, a transition at the density $ c_N= 3\pi/2$ from an
isotropic state with $Q_{ij}=0$ for $ c_0< c_N$ to a nematic state
with $Q_{ij}=S_0(n_in_j-\frac{1}{2}\delta_{ij})$, with ${\bf n}$ a
unit vector along the direction of broken symmetry, for $ c_0> c_N$.
The transition is identified with the change in sign of the
coefficient $a_2$ of $Q_{ij}$ on the right hand side of
Eq.~(\ref{st_Q}). A negative value of $a_2$ that controls the decay
rate of $Q_{ij}$ signals an instability of the isotropic homogeneous
state. A mean-field description of such a transition, which is
continuous in 2d (but first order in 3d), requires that one
incorporates cubic terms in $Q_{ij}$ in the equation for the alignment
tensor. Adding a term $-a_4 c_0^2 Q_{kl}Q_{kl}Q_{ij}$ to
Eq.~(\ref{st_Q}) we obtain $S_0 = \frac{1}{ c_0}\sqrt{-2a_2/a_4}
=\frac{1}{ c_0}\sqrt{-8(1- c_0/c_{N})/a_4}$.

If $\mu_a=0$, but $\mu_s\not=0$, there is again no stable polarized
state. The presence of a concentration of nonpolar crosslinkers does,
however, renormalize the isotropic-nematic (IN) transition, which
occurs at a lower filament density given by
\beqa \label{rhoIN} c_{IN}=\frac{ c_N}{1+\mu_s c_N/4}\;. \eeqa
The presence on nonpolar crosslinks favors filament alignment and
shifts $c_{IN}$ downward.
%%TBL
It should be noted that this occurs even with a higher effective
temperature $T_a$. 
%%TBL
A qualitatively similar result has been obtained in numerical
simulation of a two-dimensional system of rigid filaments interacting
with motor proteins grafted to a substrate~\cite{Kraikivski06}.  The
amount of nematic order $S_0$ is also enhanced by the crosslinkers,
with $S_0 =\sqrt{-2a_2/a_4c_0^2}= \sqrt{\frac{8}{a_4 c_0^2}(
  c_0/c_{IN}-1)}$.

If $\mu_a$ is finite, the system can order in both polarized and
nematic homogeneous states. The homogeneous isotropic state can become
unstable in two ways. As in the case $\mu_a=0$, a change in sign of
the coefficient $a_2$ signals the transition to a nematic (N) state at
the density $c_{IN}$ given in Eq.~(\ref{rhoIN}). In addition, the
isotropic state can become linearly unstable via the growth of
polarization fluctuations in any arbitrary direction. This occurs
above a second critical filament density,
\beqa  \label{rhoIP}c_{IP}=\frac{1}{\mu_a}\;, \eeqa
defined by the change in sign of the coefficient $a_1$ controlling the
decay of polarization fluctuations in Eq.~(\ref{st_P}). For $
c_0>c_{IP}$ the homogeneous state is polarized (P), with ${\bf
  P}\not=0$. The alignment tensor also has a nonzero mean value in the
polarized state as it is slaved to the polarization. The location of
the boundaries between the various homogeneous states is controlled by
the relative strength and concentration of active polar motor clusters
to stationary nonpolar crosslinkers. In order to simplify the
discussion we fix the value of $\mu_s$ that determined the density of
the nematic-isotropic transition to $\mu_s=0$, so that the
isotropic-nematic transition takes place at the density $c_N$ of a
suspension of rods with no crosslinkers.  One can identify two
regimes.

I) If $c_{IP}>c_N$, which corresponds to $\mu_a<1/c_N$, a region of
nematic phase exists between the isotropic and the polar state. At
sufficiently high filament and motor densities, the nematic state
becomes unstable. To see this, we linearize Eqs.~(\ref{st_P}) and
(\ref{st_Q}) by letting $Q_{ij}=Q_{ij}^0+\delta Q_{ij}$ and $\delta
P_i=P_i$. Fluctuations in the alignment tensor are uniformly stable
for $a_2<0$, but polarization fluctuations along the direction of
broken symmetry become unstable for $a_1\leq c_0 b_1 S_0/2$, i.e.,
above a critical density
\begin{equation}
c_{NP}=\frac{1}{\mu_a} \Big[ 1 + {b_1^2 \over a_4 R}\Big( 1 -
\sqrt{1 + {2 a_4 R (1-R)\over b_1^2}}\Big)\Big]
\end{equation}
where $R = c_{N} / c_{IP}$. The polarized state at $ c_0>c_{NP}$ has
\begin{eqnarray}
&&P_i^0=P_0{p}_i\;,\\
&&Q_{ij}^0=S_P({p}_i{p}_j-\delta_{ij}/2)\;,
\end{eqnarray}
with ${\bf {p}}$ a unit vector in the direction of broken symmetry
and
\begin{eqnarray}
&&P_0^2=\frac{2a_1a_2}{ c_0^2
  b_1b_2}\Big[1-\Big(\frac{2a_1}{b_1 c_0S_0}\Big)^2\Big]\;,\\
&&S_P=S_0\sqrt{1-\frac{ c_0^2
b_1b_2}{2a_1a_2}p_0^2}=2\Big|\frac{a_1}{ c_0b_1}\Big|\;.
\end{eqnarray}
%

%\begin{figure}
%\center \resizebox{0.49\textwidth}{!}{
%  \includegraphics{lmrev_fig3.eps}
%}

%\caption{ (color online) The homogeneous phase diagram for
%$g<1/4$. For all values of $\tm$ a region of nematic phase exists
%between the isotropic and polarized phases ($\gamma_P/D_r=1$,
%$g=1/10$ and $a_4=50$).}
%\label{PD1}       % Give a unique label
%\end{figure}

\begin{figure}
\center \resizebox{0.49\textwidth}{!}{
  \includegraphics{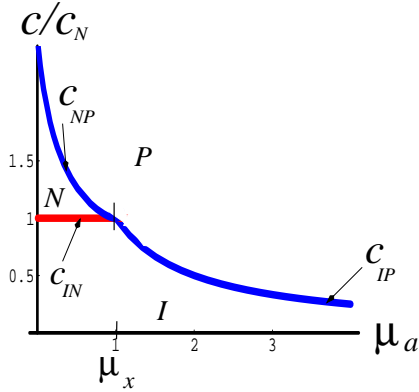}
}

\caption{ (color online) The phase diagram for $\mu_s=0$. For
$\mu_a>1/c_N$, where $c_{IN}$ and $c_{IP}$ intersect, no N state
exists and the system goes directly from the I to the P state
($\gamma_P/D_r=1$ and $a_4=50$).}
\label{PD2}       % Give a unique label
\end{figure}

II) For $\mu_a>1/c_N$, $c_{IP}>c_N$ and the polarity of motor clusters
renders the nematic state unstable at all densities and the system
goes directly from the I to the P state at $c_{IP}$, without an
intervening N state. The phase diagram has the topology shown in
Fig.~\ref{PD2}. At the onset of the polarized state the alignment
tensor is again slaved to the polarization field,
\begin{math}
Q_{ij}=\frac{b_2}{a_2} c_0~(P_iP_j-\frac{1}{2}\delta_{ij}P^2)\;,
\end{math}
and ${\bf P}=P_0{\bf {p}}$. The value of $P_0$ is determined by cubic
terms in Eq. (\ref{st_P}) not included here.

For a fixed, but nonzero value of $\mu_s$, the phase diagram has the
same topology as shown in Fig.~\ref{PD2}, but with $c_N$ replaced by
$c_{IN}$ given in Eq.~(\ref{rhoIN}). The value of $\mu_a$ where the
three phases coexist is shifted to a larger value, given by
$\mu_a=(1+\mu_sc_N/4)/ c_N$.

%The phase diagram for this case is shown in Fig.~\ref{fig:PDzerogamma1}.

Estimates of the various parameters can be obtained using a
microscopic model of the motor-filament interaction of the type
described in Appendix~\ref{sec:mappA}. Using parameter values
appropriate for kinesin ($\kappa\sim 10^{-22}\mbox{nm/rad}$
\cite{howard97}) we estimate
$\gamma_P\sim\gamma_{NP}\sim\kappa/\zeta_r=\kappa D_r/(k_BT_a)\sim
10^{-1}\mbox{sec}^{-1}$, where we used the value $D_r\sim
10^{-2}\mbox{sec}^{-1}$ appropriate for long thin rods in an aqueous
solution~\cite{Hunt93} and $T_a\sim 300\, \mbox{K}$. Using $m_a=
m_s\equiv m$, $\gamma_P=\gamma_{NP}$, $l\sim 10\mu \mbox{m}$, $b\sim
10\, \mbox{nm}$, the value of $m$ above which no nematic state exist
is found to correspond to a three-dimensional crosslinker density of
about $0.5-1\mu \mbox{M}$ and a sample thickness of order $1\mu
\mbox{m}$. This value is of order of the motor densities used in
experiments on purified microtubule-kinesin mixtures such as those of
Ref.  \cite{humphrey02}, suggesting that the filament solution in this
experiments is always in the region where the present mean field model
predicts a uniform polarized state.

On the other hand, \emph{in vitro} experiments generally fail to
observe states with uniform polarization and report the formation of
complex spatial structures. This can be understood in the context of
the hydrodynamic theory described here by examining the dynamics of
spatially varying fluctuations of the hydrodynamic fields from their
uniform value in each state. It has been shown elsewhere that such
fluctuations become unstable in a wide range of parameters. In both
isotropic and ordered states the instability arises from filament
bundling (controlled by the rate $\alpha$) that tends to build up
density inhomogeneities, eventually overtaking diffusion and driving
the formation of spatially inhomogeneous structures. This instability
is described in the next section for the isotropic state. The
instability of the nematic and polarized state is driven by the same
physical mechanisms, although the details are more subtle as in this
case one must consider the coupled dynamics of fluctuations in the
concentration and in the orientational order parameter. A complete
description can be found in Ref.~\cite{AAMCMTBL06}.

\section{Hydrodynamics of flowing active suspensions}
\label{sec:m7} In this section we display the explicit form of the
hydrodynamic equations for a suspension of active rods obtained by
coarse-graining the Smoluchowski equation, as outlined in Sections
\ref{sec:m4} and \ref{sec:m5}. Phenomenological forms of these
equations have already been used by other authors to study the
interplay of order and flow in active systems in specific
geometries~\cite{Voituriez05,zumdieck05,Muhuri06}. Our work provides a
derivation of the continuum theory starting from the dynamics of
single filaments coupled by active crosslinkers and an estimate of the
various parameters in the equations in terms of experimentally
accessible quantities. As discussed in Section \ref{sec:m4}, we limit
ourselves to the case of an incompressible suspension and neglect
inertial fluid effects. In this case the flow velocity $\bfv$ of the
suspension is determined from the solution of Stokes' equation,
\beq \label{Stokes2} \eta_0\nabla^2\bfv-\bnabla \Pi(c,{\bf P},{\bf
Q};\bm\kappa,\mu)=-\bnabla\cdot\tilde{\bm\sigma}^r(c,{\bf P},{\bf
Q};\bm\kappa,\mu);, \eeq
with the incompressibility condition
\beq \label{div_v2}\bnabla\cdot\bfv=0\;. \eeq
The pressure $\Pi$ is the sum of solvent and filament contributions,
$\Pi=\Pi_s(\rho)+\Pi_r$, and we have introduced the deviatoric stress
tensor defined by subtracting out the hydrostatic pressure,
$\Pi_r=(1/d)\sigma^r_{kk}$, as
\beq \label{sigmar_dev}
\tilde{\sigma}^r_{ij}=\sigma^r_{ij}-\delta_{ij}\Pi_r\;. \eeq
Both isotropic and ordered (polarized and nematic) suspensions will be
considered. The orientational order of the suspension affects the flow
through the dependence of the pressure $\Pi$ and the rods'
contributions to the stress tensor $\tilde{\bm\sigma}^r$ on
polarization and alignment tensor. The derivation of the constitutive
equations for these quantities was described in Section \ref{sec:m5}.

\subsection{Isotropic state}
\label{subsec:m7.1} In an isotropic suspension the only hydrodynamic
variable describing the filaments is the concentration, $c$. Its
dynamics is governed by a nonlinear convection-diffusion equation
\beqa \label{cI}\partial_t c+\bm\nabla\cdot\big({\bf
v}c\big)=\bm\nabla\cdot{\cal D}(c)\bm\nabla c\;, \eeqa
where ${\cal D}(c)$ is an effective (concentration-dependent)
diffusion coefficient, softened by active processes. It is given
by
\beq \label{Dciso}{\cal D}(c)=\frac{3D}{4}(1+v_0c)-\alpha \tm_a
c\;, \eeq
with $\tm_a=m_a b^2$. The first term on the right hand side of
Eq.~(\ref{Dciso}) is the diffusion coefficient of long thin rods, with
$D=D_\parallel=2D_\perp$, including excluded volume corrections, with
$v_0=2 l ^2/\pi$.  The second term on the right hand side of
Eq.~(\ref{Dciso}) arises from filament bundling driven at the rate
$\alpha$ given in Eq.~(\ref{alpha_par}) and promotes density
inhomogeneities. Equation (\ref{cI}) for the concentration couples to
the Stokes equation, Eq.~(\ref{Stokes2}), with
\beq
\label{Pir_ISO}\Pi_r^{I}(c,\mu)=k_BT_ac\Big(1+\frac{2c}{\pi}\Big)
+\tm_a\alpha \frac{5k_BT_a}{432D}c^2\;, \eeq
and
\beq \label{sigma_ISO} \tsigma^{r,I}_{ij}=\Big(2\eta_0
+\frac{k_BT_a}{96D}c\Big)u_{ij}\;. \eeq
In an isotropic active suspension there are no active contributions to
the deviatoric part of the stress tensor, which has the form usual for
passive rods~\cite{DoiEdwards}. There is, however, an active
contribution to the pressure corresponding to the second term on the
right hand side of Eq.~(\ref{Pir_ISO}). The first term on the right
hand side of Eq.~(\ref{Pir_ISO}) is standard for passive rods.

\begin{figure}
\center \resizebox{0.7\textwidth}{!}{%
 \includegraphics{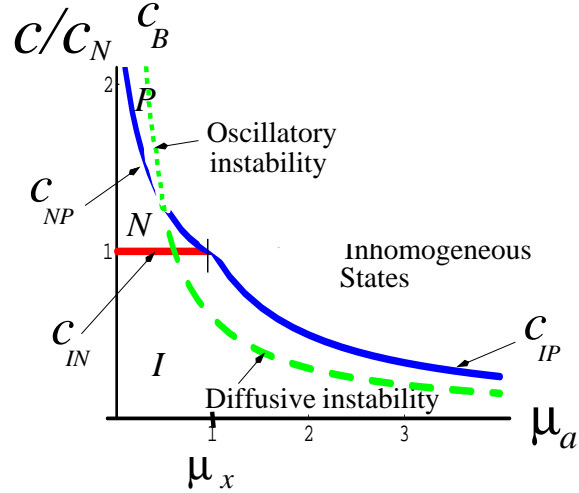}
 }

\caption{(color online) The phase diagram of homogeneous states
for $\mu_s=0$ in the plane of filament density, $c_0$, and motor
activity $\mu_a$, as defined in Eq.~(\ref{muA}), showing the
location of the bundling instability at $c_0=c_B$. The horizontal
line at $c_0=c_N$ for the isotropic-nematic transition crosses
$c_{IP}$ at $\mu_a\mu_x=1/c_N$.  The $c_B$ line may lie above the
$c_{NP}-c_{IP}$ line or cross through the {\em N} and {\em I}
states, as shown in the figure ($ l \gamma_P/\alpha=0.1$,
$a_4=50$), depending on the value of $ l \gamma_P/\alpha$, a
numerical parameter to leading order independent of ATP
consumption rate. The instability of the {\em I} and {\em N}
states is diffusive (dashed line), while the instability of the
{\em P} state is oscillatory (dotted line).}
\label{Fig:rho_instability}
\end{figure}

The homogeneous isotropic state in a quiescent suspension is
characterized by $\bfv=0$ and $c=c_0$. As discussed in the
literature~\cite{kruse00,tblmcm03,AAMCMTBL06}, the homogeneous state
becomes unstable at high filament and motor concentration due to
contractile effects generated by motor-induced filament bundling.
Bundling is the main mechanism responsible for the instability of both
isotropic and ordered homogeneous states in quiescent suspensions. It
is therefore instructive to display explicitly the details of this
instability for the simple isotropic case. The examine the dynamics of
fluctuations in the isotropic state we let $c=c_0+\delta c$ and ${\bf
  v}=\delta \bfv$ in Eq.~(\ref{cI}) and only keep terms of first order
in the fluctuations. Incompressibility requires
$\bnabla\cdot\delta\bfv=0$ and the linearized equation for $\delta c$
is simply
\beq
\partial_t\delta c={\cal D}(c_0)\nabla^2\delta c\;.
\eeq
Expanding $\delta c$ in Fourier components, $\delta c(\bfr,
t)=\sum_{\bf k}=c_{\bf k}(t)e^{i{\bf k}\cdot\bfr}$, one finds
immediately that the relaxation of the Fourier amplitudes, $c_{\bf
  k}(t)=c_{\bf k}e^{-z_c(k)t}$, is controlled by a diffusive mode
\beq z_c(k)={\cal D}(c_0)k^2\;. \eeq
Density fluctuations become unstable when $z_c(k)<0$,
corresponding to ${\cal D}(c_0)<0$ or $c>c_B$, where
\beq\label{cB}
c_B=\frac{3D}{4\tm\alpha-3Dv_0}\sim\frac{3D}{4\tm\alpha}\eeq
is the concentration above which bundling overtakes diffusion.
Using $\alpha\sim (b/ l )u_0$, we can express the density $c_B$
in terms of the activity parameter $\mu_a$ defined in
Eq.~(\ref{muA}) as $c_B=\frac{9}{2\mu_a}( l \gamma_P/\alpha)$,
where we have used $D_r=D/(6 l ^2)$. A possible location of this
instability line in the phase diagram is shown in
Fig.~\ref{Fig:rho_instability}.

\subsection{Nematic State}
\label{subsec:m7.2} The continuum variables describing the large-scale
dynamics of an active nematic solution are the density and flow
velocity of the solution and the concentration and alignment tensor of
the filaments.  For simplicity we consider only the case where there
are no stationary apolar cross-linkers, i.e., $m_s=0$. In this case
the transition from the isotropic to the nematic state occurs at the
value $c_N$ of passive suspensions. A finite fraction of stationary
apolar crosslinkers lowers the value of the transition density, as
discussed in Section \ref{sec:m6}. In addition, it tends to stiffen
all the liquid crystal elastic constants~\cite{AAMCMTBL06}. In the
absence of external forces, the equations for filament concentration
and alignment tensor are given by
\beqa \label{cN}\Big(\partial_t +{\bf
v}\cdot\bm\nabla\Big)c=\partial_i{\cal
D}_{ij}\partial_jc+\partial_i{\cal D}^Q\partial_j(cQ_{ij})\;,
\eeqa
\beqa \label{QijN} \Big(\partial_t +{\bf
v}\cdot\bm\nabla\Big)(cQ_{ij})&=&cF_{ij}(\bm\kappa,{\bf Q})
 +H_{ij}(c,{\bf Q})\;. \eeqa
The tensor $F_{ij}$ describes anisotropic convective and flow
alignment effects and has the familiar form for passive nematic
liquid crystals, as given in Eq.~(\ref{Fij}). At low density with
the closure approximation described in Ref.~\cite{AAMCMTBL06} the
alignment parameter has the value $\lambda=1/(2S_0)$, with $S_0$
the nematic order parameter defined in Eq.~(\ref{Q_def_macro}).
The tensor $H_{ij}$ plays the role of the equilibrium molecular
field for passive nematic liquid crystals, but it contains various
active corrections. It is given by
%\begin{widetext}
%
\beqa \label{H} H_{ij}(c,{\bf Q})&\simeq&K\nabla^2(c Q_{ij})
+K'\big[\partial_i\partial_k(cQ_{jk})+\partial_j\partial_k(cQ_{ik})
-\delta_{ij}\partial_k\partial_l(cQ_{kl})\big]\nonumber\\
&&+\partial_k\big(K_{ijkl}\partial_lc\big)
-4D_r\Big(1-\frac{c}{c_{N}}\Big)cQ_{ij}-D_ra_4c^3Q_{kl}Q_{kl}Q_{ij}\;,
\eeqa
%
%\end{widetext}
where
\beqa \label{Dij_N}&&{\cal D}_{ij}(c,{\bf
Q})=\frac{3D}{4}\Big[1+\Big(1-\frac{2}{3}S^2\Big)\frac{3c}{c_N}-\frac{4\alpha\tm_a
c}{3D}\Big]\delta_{ij} \nonumber\\
&&\hspace{0.6truein} +\Big(\frac{Dv_0}{2}-\frac{4}{3}\alpha\tm_a
\Big)cQ_{ij}\;,\\
\label{DQ}&&{\cal D}^Q(c)=\frac{D}{2}\Big(1-\frac{c}{c_N}\Big)-\frac{2\alpha\tm_a c}{3}\;,\\
&&K_{ijkl}(c)=\Big[\frac{D}{16}(1+v_0c)-\frac{2}{3}\alpha\tm_a
c\Big]
%\nonumber\\&&\hspace{0.5in}\times
(\delta_{ik}\delta_{jl}+\delta_{jk}\delta_{il}-\delta_{ij}\delta_{kl})\;,\\
\label{K}&&K(c)=\frac{7D}{12}\Big(1-\frac{c}{c_N}\Big)\;,\\
\label{Kp}&&K'(c)=\frac{D}{6}\Big(1-\frac{c}{c_N}\Big)-\frac{\alpha\tm_a
c}{18}\;. \eeqa
The last term on the right hand side of Eq.~(\ref{H}), with $a_4>0$
has been introduced by hand. It arises from a quartic terms in the
free energy of an equilibrium nematic and determines the magnitude of
orientational order in passive rod solutions.  It is apparent from the
form of the various elastic constants in Eqs.~(\ref{Dij_N}-\ref{Kp})
that bundling (described by the parameter $\alpha$ of
Eq.~(\ref{alpha_par})) always decreases the elastic constants of the
nematic and therefore ultimately renders the uniform ordered state
unstable.

The flow velocity of the suspension is again obtained from Stokes'
equation, Eq.~(\ref{Stokes2}), with a rods' contribution to the
pressure given by
\beqa \label{pressuref_N}
\Pi_r^N(c,\mu)&=&k_BT_ac\Big[1+\frac{3 c}{2 c_N}\Big(1-\frac{2}{3}S^2\Big)\Big]
+k_BT_a\frac{c}{36D}u_{kl}Q_{kl}\nonumber\\
&&+\tm_a\alpha \frac{k_BT_a}{432D}c^2\big(5-S^2\big) \;, \eeqa
where the last term is new and arises from activity. The filament
contribution to the deviatoric stress tensor is given by
\begin{eqnarray}
\tsigma_{ij}^{r,N} &=&  2 k_B T_ac \Big(1 - {c \over c_{N}} \Big)
Q_{ij} +\tm_a\alpha \frac{8k_B T_a}{432D} c^2 Q_{ij}\nonumber\\&&
+k_BT_a\frac{c}{24D}\Big[\frac{1}{2}u_{ij}
+\frac{2}{3}\Big(u_{ik}Q_{kj}+u_{jk}Q_{ki}-\delta_{ij}
u_{kl}Q_{kl}\Big)\Big)]\;. \label{act_stress_N} \eeqa
Activity modifies the stress tensor of a nematic in two ways. The
first term on the right hand side of Eq.~(\ref{act_stress_N}) is
equilibrium-like, in the sense that it can be obtained from the
corresponding term in the stress tensor of passive rods,
$\tsigma_{ij}^{r,{\rm passive}}=2 k_B T \Big(1 - {c \over c_{N}}
\Big) Q_{ij}$ by letting $T\rightarrow T_a$ (and replacing the
transition density $c_N$ by $c_{IN}$, when $m_s\not=0$). The
second term on the right hand side of Eq.~(\ref{act_stress_N}) is
a truly nonequilibrium contribution. It was first proposed
phenomenologically by Hatwalne and collaborators~\cite{Hatwalne04}
who argued that an active element in solution behaves like a force
dipole.  Correlations among the axis of each dipole build up
orientational order and yield active contributions to the stress
tensor proportional to the orientational order parameter,
$Q_{ij}$. Our microscopic derivation~\cite{TBLMCMPRL06} yields an
estimate for the coefficient of this term (undetermined, even in
sign, in the phenomenological theory) and shows that the active
cross-linkers yield contractile stresses ($\alpha>0$). Finally,
the third term on the right hand side of Eq.~(\ref{act_stress_N})
is the viscous contribution which has the standard form for a
solution of rod-like filaments. Finally we note that active
contributions proportional to the parameter $\beta$ given in
Eq.~(\ref{beta_par}) do not appear in the hydrodynamics of the
nematic phase. This is expected as terms proportional to $\beta$
break the inversion symmetry of the ordered state and can only
appear in a system with polar order.

\subsection{Polarized State}
\label{subsec:m7.3} The coarse-grained variables describing the
dynamics of an active polarized suspension are the density and
flow velocity of the solution and the concentration and
polarization of the filaments.
%Although the polarization vector
%itself is not a true hydrodynamic variable as only the
%\emph{direction} of polarization, not its magnitude, is associated
%with a broken symmetry in the ordered state, we will first need to
%consider the full equation for the polarization to discuss the
%possible steady states of the suspension in the presence of an
%imposed flow.
As shown in Section \ref{sec:m6}, in a polarized state the
alignment tensor is slaved to the polarization field and it is not
an independent continuum field. On the other hand, since our
theory only considers terms that are quadratic in the fields,  a
nonzero value for $|{\bf P}|$  is only obtained by considering the
coupled equations for ${\bf P}$ to $Q_{ij}$ and eliminating
$Q_{ij}$ in favor of ${\bf P}$ in the polarization equation to
generate a term of order $({\bf P})^3$.  To see, consider a
filament density well into the polarized state, with $c>c_{IN}$
and $c>c_{IP}$, so that both coefficients $a_1=1-c/c_{IP}$ and
$a_2=1-c/c_{IN}$ in Eqs.~(\ref{st_P}) and (\ref{st_Q}) satisfy
$a_1<0$ and $a_2<0$. Setting the left hand side of
Eqs.~(\ref{st_P}) and (\ref{st_Q}) to zero, we solve
Eq.~(\ref{st_Q}) for $Q_{ij}$ to obtain
\beq\label{Qij_sol}
Q_{ij}=\frac{b_2c}{a_2}\Big(P_iP_j-\frac{1}{2}\delta_{ij}P^2\Big)\;.
\eeq
This solution, substituted in Eq.~(\ref{st_P}), yields a term
$\sim P^2 P_i$ on the right hand side of Eq.~(\ref{st_P}) which
has solution $P^2=(2a_1a_2)/(b_1b_2c^2)$.

The continuum equations for the polarized state are obtained by
assuming that the alignment tensor relaxes on microscopic time
scales to the form given by Eq.~(\ref{Qij_sol}), which is then
used to eliminate $Q_{ij}$ in favor of ${\bf P}$. With the
exception of homogeneous terms, such as the ${\cal O}(({\bf
  P})^3)$ term just described, this  leads to a high density
renormalization of the various coefficients in the continuum
equations, but does not generate any new terms. For the sake of
simplicity in the following we neglect this renormalization and
only keep those terms in the polarization equation generated by
the coupling to the alignment tensor that have a qualitatively new
structure. We also neglect all excluded volume corrections. The
equation for filament concentration is given by
\beqa \label{cP}\partial_t c&=&-\bm\nabla\cdot c\big({\bf
v}-\frac{7}{36}\tm_a\beta c{\bf P}\big)+\partial_i\big({\cal
D}_{ij}^p(c)\partial_jc\big)\nonumber\\
&&-\frac{1}{2}\alpha\tm_a\partial_i\big[c^2\partial_j(P_iP_j)\big]\;,
\eeqa
with
\beq \label{Dpij}{\cal D}_{ij}^p(c,{\bf
P})=\Big(\frac{3D}{4}-\alpha\tm_a c\Big)\delta_{ij}-\alpha\tm_a
c\Big(P_iP_j+\frac{1}{2}\delta_{ij}P^2\Big)\;. \eeq
The equation for the polarization vector has the form
\beqa \big(\partial_t+\bfv\cdot\bnabla\big)(c{\bf
P})&=&\frac{1}{2}(\bnabla\times\bfv)\times(c{\bf
P})+\frac{\lambda_P}{2}\big[\bnabla\bfv+(\bnabla\bfv)^T\big]\cdot
c{\bf P}\nonumber\\
&&+{\bf H}(c,{\bf P})\;.\label{Peqn} \eeqa
where ${\bf H}(c,{\bf P})$ generalizes the molecular field of
equilibrium polar fluids~\cite{WKMCMKS06} by including active
contributions. It is given by
\beqa \label{Hi}H_i(c,{\bf P})&\simeq&-\big[D_r-\gamma_P\tm_a
c+a_3P^2\big]cP_i+\frac{2}{9}\tm_a\beta c\partial_ic
-\frac{1}{36}\tm_a\beta\partial_j\big[c^2\big(P_iP_j-\frac{5}{2}\delta_{ij}P^2\big)\big]\nonumber\\
&&+\big[\partial_jK_p\partial_i(cP_j)+\partial_iK_p\partial_j(cP_j)\big]+\partial_jK_p\partial_j(cP_i)\nonumber\\
&&-\partial_j{\cal D}^p_{ijk}(c,{\bf P})\partial_kc
+\gamma_P\frac{\tm_a c}{24}\nabla^2(cP_i)\;,
 \eeqa
where
\beqa
&&K_p(c)=\frac{D}{8}-\frac{\alpha\tm_a}{4}c\;,\\
&&K'_p(c)=\frac{5D}{8}-\frac{\alpha\tm_a}{4}c\;, \eeqa

 \beqa\label{DPijk} {\cal
D}^p_{ijk}(c,{\bf P})=c\Big[\Big(\frac{Dv_0}{8}+\frac{\alpha
\tm_a}{3}\Big)(P_i\delta_{jk}+\delta_{ij}P_k)
+\Big(\frac{17Dv_0}{8}+\frac{2\alpha
\tm_a}{3}\Big)P_j\delta_{ik}\Big]\;. \eeqa
The parameter $a_3$  determines the value $P_0$ of the magnitude
of the polarization in a quiescent (active) suspension, with
$P_0^2=a_3/[D_r(c/c_{IP}-1)]$.

In contrast to the case of the nematic, all three active
mechanisms of motor-induced filament dynamics controlled by
$\alpha$, $\beta$ and $\gamma_P$ appear in the hydrodynamic
equations of polarized active suspension. Polarization sorting at
a rate $\beta\sim u_0$ yields novel convective contributions in
the first term on the right hand side of the equation for the
filament concentration, Eq.~(\ref{cP}). In an equilibrium
suspensions the filament concentration is convected with the
suspension flow velocity, $\bfv$. In an active polar suspension,
in contrast, the filament concentration is convected with the
effective velocity $\sim\bfv+\tm_a\beta{\bf P}$. The terms  linear
in the gradients proportional to $\beta$  in the polarization
equation are of similar origin. These terms were also incorporated
in the continuum description of self-propelled particles proposed
by Simha and Ramaswamy. Bundling effects controlled by the rate
$\alpha$ soften both the diffusion constant ${\cal
D}_{ij}^p(c,{\bf P})$ in the concentration equation and the
effective bend and splay elastic constants $K_p$ and $K'_p$ of the
polar fluid. Finally, the rotation rate $\gamma_P$ builds up polar
order and controls the very existence of a polar state.

For an incompressible suspension, the flow field $\bfv$ is
obtained again from the Stokes equation, Eq.~(\ref{Stokes2}). The
filament contribution to the pressure is given by
\beqa \label{Pi_rP} \Pi_r^P=k_BT_a
c\Big(1+\frac{c}{\pi}\Big)+\tm_a\alpha
\frac{k_BT_a}{144D}c^2\Big(\frac{5}{3}+2P^2\Big)\;. \eeqa
The filament contribution to the deviatoric stress tensor of a
polarized suspension is
\begin{eqnarray}\label{sigmaP}
\tilde{\sigma}_{ij}^{r,P} & =& \tm_a\alpha \frac{k_B T_a}{72D}
c^2\Big( P_i P_j - {1 \over 2} \delta_{ij}P^2 \Big)+{k_BT_a\over
48D}c u_{ij}
\nonumber\\
&& +\frac{k_BT_a}{36D} \frac{c^2b_2}{a_2}\Big[
u_{ik}P_kP_j+u_{jk}P_kP_i-\delta_{ij}P_ku_{kl}P_l-u_{ij}P^2\Big]\nonumber\\
&&+\tm_a\beta\frac{k_B T_a}{432D} c^2\Big[\partial_j
P_i-\frac{1}{2}\delta_{ij}\bnabla\cdot{\bf
P}-\frac{1}{4}\Big(\partial_i P_j-\partial_j P_i\Big)\Big]. 
\end{eqnarray}
The first term on the right hand side of Eq.~(\ref{sigmaP}) is the
active contribution to the stress tensor first discussed by
Hatwalne and collaborators for a nematic
suspension~\cite{Hatwalne04}.  The second and third term arise
from the viscous coupling of filaments to the solvent. Finally the
last term contains active contributions proportional to gradients
of the polarization. These are controlled by the polarization
sorting rate $\beta\sim u_0$. Terms of these type are unique to
the polarized state and vanish in a nematic suspension. They are
expects to play an important role in renormalizing the rate of
stress relaxation via reptation.

Continuum equations for a polarized active suspension have been
written down phenomenologically by several
authors~\cite{Simha02,Voituriez06,Muhuri06}. It is useful to make
contact with this work. The phenomenological description can be
recovered from our model by making a few simplifying
approximations. An equation for the concentration of filaments of
the form given in Eq.~(\ref{cP}) was proposed by Ramaswamy and
collaborators~\cite{Simha02,Muhuri06}, although these authors
neglected the diffusion terms, which play a crucial role in
controlling the bundling instability of quiescent suspensions. The
equation for the polarization vector ${\bf P}$ reduces to the form
used by Voituriez et al.~\cite{Voituriez06} and by Simha et
al.~\cite{Simha02,Muhuri06}, if all terms containing higher order
gradients of the concentration ($P_i\nabla^2c$, $P_i(\bnabla
c)^2$, ${\bf P}\cdot\bnabla\partial_ic$, $({\bf P}\cdot\bnabla
c)(\partial_ic)$), as well as terms containing both gradients of
concentration and of polarization ($(\bnabla\cdot{\bf
  P})(\partial_i c)$, $(\partial_j c)(\partial_jP_i)$, $(\partial_j
c)(\partial_iP_j)$) are neglected. With this approximation
Eq.~(\ref{Peqn}) becomes
\beqa\label{Peqn_simple}
\big(\partial_t+\bfv\cdot\bnabla\big)P_i=&&\Gamma\Big(1-\frac{|{\bf
P}|^2}{P_0^2}\Big)P_i
+\frac{1+\lambda_P}{2}(\partial_jv_i)P_j-\frac{1-\lambda_P}{2}(\partial_iv_j)P_j\nonumber\\
&&-w_1c({\bf P}\cdot\bnabla)P_i-w_2cP_i(\bnabla\cdot {\bf
P})+w_3c\partial_i|{ \bf P}|^2\nonumber\\
&&+\Big[\Big(w_4+w_5P^2\Big)\delta_{ij}-w_6P_iP_j\Big]\partial_jc\nonumber\\
&& +(K_1-K_3)\partial_i\bnabla\cdot{\bf P}+K_3\nabla^2P_i\;,\eeqa
where $\Gamma=\gamma_P\tm_a c-D_r>0$ and $P_0^2=\Gamma/a_3$.  Here the coefficients $w_i$ have the form $w_i=c_i\tm_a\beta$, with
$c_i$ numerical coefficients of order one. Note, however,  that the terms proportional to $w_i$ other than $w_1$ are equilibrium-like, in the sense that they could also be obtained from a polar contribution to the free energy of the form $\delta F_p=\int_{\bfr}\Big[B_1\delta c(\bm\nabla\cdot{\bf P})+B_2c|{\bf P}|^2(\bm\nabla\cdot{\bf P})+B_3|{\bf P}|^2{\bf P}\cdot\bm\nabla c+...\Big]$. The $w_1$ term, in contrast,
is a true nonequilibrium contribution induced by activity and cannot be obtained from a free energy. 
All the remainder $w_i$'s  contain in general both equilibrium-like  contributions determined by the $B_i$ and nonequilibrium ones proportional to $\beta\sim R_{ATP}$. Finally, $K_1$ and $K_3$ are
the splay and bend elastic moduli, respectively, with
\beqa
&&K_1(c)=\frac{7D}{8}-\frac{3\tm_a\alpha c}{4}+\frac{\tm_a\gamma_P c}{24}\;,\label{K1}\\
&&K_3(c)=\frac{5D}{8}-\frac{\tm_a\alpha
c}{4}+\frac{\tm_a\gamma_Pc}{24}\label{K3}\;. \eeqa
The first term on the right hand side of Eq.~(\ref{Peqn_simple})
guarantees the formation of a uniformly polarized state with
$|{\bf P}|=P_0$. The next two terms are conventional couplings of
liquid crystalline order and flow, with $\lambda_P$ the flow
alignment parameter. Our low density calculation yields
$\lambda_P=1/2$. The three terms on the second line are
nonequilibrium terms analogous to those first written down by
Toner and Tu in models of
flocking~\cite{TonerTu95,TonerTu98,Toner05}. The third line
describes nonequilibrium changes in polarization driven by
concentration gradients. Only the first of these terms ($\sim
w_4$) is generally included in phenomenological theories.
Equations~(\ref{K1}) and (\ref{K3}) show that motor-induced
filament bundling ($\sim\alpha$) softens both the splay and bend
elastic constants, while polar crosslinkers ($\sim\gamma_P$) tend
to stiffen them.
%%TBL
Such effects, i.e. the dependence of elastic constants on the active elements are clearly beyond the scope of phenomenological theories with arbitrary elastic constants.
In addition, the microscopic derivation also provides contributions to
the stress tensor which are higher order in gradients without the need for new unknown parameters, e.g. the expression for the stress tensor given in
Eq.~(\ref{sigmaP}) contains novel contributions proportional to
gradients of polarization that were not considered by other
authors~\cite{Simha02,Hatwalne04,kruseEPJ06,Voituriez05} but whose microscopic origin is the same as those of lower order in the gradients.
%%TBL

\vspace{0.2in}

\noindent{\bf Acknowledgements}\\
Some of the work described in this article has been done in
collaboration with Aphrodite Ahmadi. MCM acknowledges  support
from the National Science Foundation through grants DMR-0305407
and DMR-0219292. TBL acknowledges support from the Royal Society.
We also thank S. Ramaswamy for many helpful discussions.

\appendix
\section{Appendix: Derivation of the rods' stress tensor}
\label{sec:mappA}

We model very long, thin rods as  rigid strings of spherical beads
of diameter $b \ll l$ suspended in a fluid of viscosity $\eta_0$. We
assume each rod consists of an odd number $l/b=2M+1$ of such
beads, a sketched in Fig.~\ref{Fig:kebab}. The beads on the
$\alpha$-th rod are indexed by an integer $m$ that runs from $-M$
to $+M$ and the center of the $m$-th bead is at
$\bfr_{\alpha}(m)=\bfr_\alpha+mb\uvec_\alpha$, with
$\bfr_\alpha=\sum_m\bfr_\alpha(m)$ the center of mass of the rod.
\begin{figure}
\center \resizebox{0.35\textwidth}{!}{
  \includegraphics{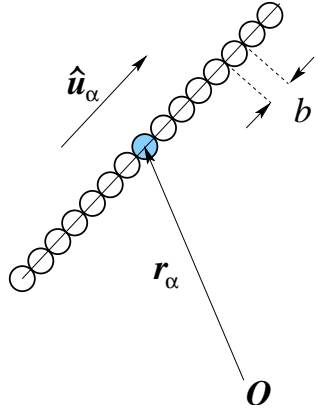}}

\caption{(color online) The bead model of a rigid filament.}
\label{Fig:kebab}       % Give a unique label
\end{figure}
Momentum conservation at low Reynolds number is described by the
Stokes equation
\beq \eta_0 \bnabla^2 \bfv (\bfr) - \bnabla \Pi - \sum_\alpha
\sum_{m=-M}^{M} \delta \left(\bfr - \bfr_\alpha(m) \right) {\bf
f}^h_\alpha(m) =0\;, \eeq
where we have modeled each bead as a point-force on the fluid at
the position of its center of mass. Fax\'en's theorem for the
hydrodynamic force on a sphere in an inhomogeneous flow relates
the force ${\bf f}^h_\alpha(m)$ exerted by the fluid on a bead to
the flow velocity field $\bfv_0(\bfr)$ at the bead's position in
the absence of that bead, according to
\beq \label{Faxen} {\bf
f}^h_\alpha(m)=-\zeta_b\big[\bfv_\alpha(m)-\bfv_0(\bfr_\alpha(m))\big]\;,
\eeq
where
$\bfv_\alpha(m)=\bfv_\alpha+mb\bm\omega_\alpha\times\uvec_\alpha$
is the velocity of the bead, with $\bfv_\alpha$ and
$\bm\omega_\alpha$ the center of mass and angular velocity of the
rod, and $\zeta_b=3 \pi \eta_0 b$ is the Stokes friction
coefficient of a sphere of diameter $b$ in an unbounded fluid of
viscosity $\eta_0$. Using the linearity of the Stokes equation and
the principle of superposition, the velocity of fluid at the
position of the bead is given by
\beq \bfv_0(\bfr_\alpha(m))=\bfv(\bfr_\alpha(m)) - \sum_{n\ne m}
{\bf H} \left(\bfr_\alpha(m)-\bfr_\alpha(n)\right) \cdot
\bff^h_\alpha(n)\;, \eeq
where $\bfv(\bfr)$ is the velocity of the fluid taking account of
the presence of other rods and $\ds H_{ij}(\bfr)= {1 \over 8 \pi
\eta_0 r}\left( \delta_{ij}+ \hat{r}_i\hat{r}_j \right)$ is the
Oseen tensor. Here the hydrodynamic interactions between beads on
the same rod have been included explicitly, while the hydrodynamic
coupling to other rods is implicitly taken into account in
determining the flow velocity $\bfv(\bfr)$. The force on bead $m$
on the $\alpha-$th rod is therefore given by
\beq \bff^h_\alpha(m) = - \zeta_b
\left[\bfv_\alpha(m)-\bfv\left(\bfr_\alpha(m)\right) \right] - {3
\over 8} \sum_{n\ne m} {1 \over |n-m|} (\bm\delta +
\uvec_\alpha\uvec_\alpha)\cdot \bff^h_\alpha(n)\;. \eeq
Now we
take the limit $l \gg b$ and introduce the continuous variable
$s=bm$, where $-l/2 \le s \le l/2$, so that  $\bfr_\alpha(s) =
\bfr_\alpha + s \uvec_\alpha$. The hydrodynamic force per unit
length, $ {\cal F}^h_\alpha(s)$, satisfies the equation
\beq {\cal
F}^h_\alpha(s) = - \zeta_s
\left(\bfv_\alpha(s)-\bfv\left(\bfr_\alpha(s)\right) \right) - {3
\over 8} \int_{|s-s'|\ge b}  {ds' \over |s-s'|} (\bm\delta +
\uvec_\alpha\uvec_\alpha)\cdot {\cal F}^h_\alpha(s') \; , \eeq
where $\zeta_s = 3 \pi \eta_0 = \zeta_b/b$. Approximating
\beq {1 \over |s-s'|}\approx\left\langle {1 \over
|s-s'|}\right\rangle'\delta (s-s')\;,\eeq
where
\beq\ds \left\langle {1 \over |s-s'|}\right\rangle' = {1 \over
l^2}\int_{s}\int_{s'} \Theta(|s-s'|-b){1 \over |s-s'|} = {2 \ln
(l/2b)}\;, \eeq
with $\Theta(x)$ the Heaviside function, we obtain
\beq {3 \ln (l/2b) \over 4 }(\bm\delta +
\uvec_\alpha\uvec_\alpha)\cdot {\cal F}^h_\alpha(s) \simeq -
\zeta_s \left[\bfv_\alpha(s)-\bfv\left(\bfr_\alpha(s)\right)
\right]\;. \label{eq:segment_vel} \eeq
Since $\bfv_\alpha(s) = \bfv_\alpha + s \bm\omega_\alpha \times
\uvec_\alpha$, then integrating equation (\ref{eq:segment_vel})
over $s$, we can obtain expressions for the hydrodynamic force,
$\bfF_\alpha^h = \langle{\cal F}^h_\alpha(s)\rangle_s$ and torque
$\bm\tau_\alpha^h= \langle \uvec_\alpha s \times {\cal
  F}^h_\alpha(s)\rangle_s$ at the center of mass of a rod, with
$\langle ...\rangle_s=\int_{- l /2}^{ l /2} ds...$ as
\beqa
-\bm\zeta^{-1}(\uvec_\alpha) \cdot \bfF_\alpha^h &=&
\bfv_\alpha - {1 \over l} \langle\bfv(\bfr_\alpha(s))\rangle_s\;,  \\
-{1 \over \zeta_r} \bm\tau_\alpha^h &=& \bm\omega_\alpha -
I^{-1}{1 \over l} \langle \uvec_\alpha \times s
\bfv(\bfr_\alpha(s))\rangle_s\;, \eeqa
where $\zeta_{ij}(\uvec)=\zeta_\perp
(\delta_{ij}-\uhat_i\uhat_j) + \zeta_{\|}\uhat_i\uhat_j$,
$\zeta_\perp = 2\zeta_{\|}={ 4 \pi \eta_0 l \over \ln (l/2b)}$,
$\zeta_r = {\pi l^3 \eta_0 \over 3 \ln (l/2b)}$ and $I= l^2/12$.
Performing a Taylor expansion of the fluid velocity about the
center of mass, we obtain to lowest order in gradients
\beqa
-\bm\zeta^{-1}(\uvec_\alpha) \cdot \bfF_\alpha^h &=&
\bfv_\alpha - \bfv(\bfr_\alpha) - {I \over 2}
\left( \uvec_\alpha \cdot \bm\nabla \right)^2 \bfv (\bfr_\alpha) + O(\nabla^4)\;,\label{vgrad}\\
-{1 \over \zeta_r} \bm\tau_\alpha^h &=& \bm\omega_\alpha -
\uvec_\alpha \times (\uvec_\alpha \cdot \bm \nabla)
\bfv(\bfr_\alpha) + O(\nabla^3)\;. \eeqa
Finally, we require that the hydrodynamic forces and torques be
balanced  by all other forces and torques on the rod. This gives
\beqa
\bfF^h_\alpha &=& \bm\nabla_\alpha U_{ex} + k_B T_a \bm\nabla \ln \hat{c} - \bff_\alpha^a\;, \\
\bm\tau_\alpha^h &=& {\cal R}_\alpha U_{ex} + k_B T_a {\cal
R}_\alpha \ln \hat{c} - \bm\tau_\alpha^a\;, \eeqa
where we have
included contributions from fluctuations (non-equilibrium osmotic
pressure), excluded volume interactions and active driving by the
motors.
Using Eqs.~(\ref{eq:segment_vel}) and  (\ref{vgrad}), we can
calculate the hydrodynamic force per unit length on the rod as
\beqa {\cal F}^h_{\alpha i}(s) &=& \zeta_{ij}(\uvec_\alpha) \big[
v_{\alpha j} -v_j(\bfr_\alpha) + s \left((\bm\omega_\alpha \times
\uvec_\alpha)_j - (\uvec \cdot \bm\nabla_\alpha)
v_j(\bfr_\alpha)\right) \nonumber\\
&&- {s^2 \over 2} (\uvec \cdot \bm\nabla_\alpha)^2
v_j(\bfr_\alpha)\big]\;. \eeqa
from which we obtain Eq.~(\ref{eq:stress_ftau}).
Furthermore, defining the translational and rotational currents as
\beqa \bfJ_c (\bfr,t) &=&  \left\langle \sum_{\alpha}\bfv_\alpha \delta(\bfr-\bfr_\alpha(t)) \right\rangle \\
{\cal J}_c (\bfr,t) &=&  \left\langle\sum_{\alpha}
\bm\omega_\alpha \delta(\bfr-\bfr_\alpha(t)) \right\rangle \; ,
\eeqa
the Smoluchowski equation (\ref{smoluchowski}) for the dynamics of
$\hat{c}(\bfr,\uvec,t)$ follows.

From equation (\ref{eq:stress_ftau}), we perform the
coarse-graining procedure to obtain the stress tensor. Retaining
all terms of first order in gradients of the hydrodynamic fields,
the pressure is
\beqa \label{Pi_P} \Pi_r^P=k_BT_a
c\Big(1+\frac{c}{\pi}\Big)+\tm_a\alpha
\frac{k_BT_a}{144D}c^2\Big(\frac{5}{3}+2P^2\Big)\;, \eeqa
and the deviatoric stress tensor is given by
\begin{eqnarray}
\tilde{\sigma}_{ij}^A & =& 2 k_B T_a c\Big(1 - {c \over c_{IN}}
\Big)  Q_{ij}
-k_B T_a { c^2 \over c_{IP}} \Big( P_i P_j - {1 \over 2}  \delta_{ij}P^2 \Big)\nonumber \\
&&+  \tm_a\alpha \frac{k_B T_a}{72D} c^2\Big({4  \over 3} Q_{ij} +
P_i
P_j - {1 \over 2}\delta_{ij}   P^2 \Big)\nonumber\\
&&+\tm_a\beta \frac{2k_BT_a}{432D}c^2\Big[\partial_j
P_i-\frac{1}{2}\delta_{ij}\bnabla\cdot{\bf
P}-\frac{1}{4}\Big(\partial_i
P_j-\partial_j P_i\Big)\nonumber\\
&&+\frac{5}{3}\Big(Q_{jk}\partial_kP_i-P_i\partial_kQ_{jk}
-Q_{ik}(\partial_jP_k+\partial_kP_j)+
(P_k\partial_j+P_j\partial_k)Q_{ik}\nonumber\\
&&-Q_{ij}\bnabla\cdot{\bf P}+{\bf P}\cdot\bnabla Q_{ij}\Big)
+\frac{2}{3}\Big(Q_{jk}(\partial_kP_i+\partial_iP_k)-(P_i\partial_k+P_k\partial_i)Q_{jk}\Big)\nonumber\\
&&+\frac{5}{6}\delta_{ij}\Big(Q_{kl}\partial_kP_l-P_l\partial_kQ_{kl}\Big)\Big]\;. 
\label{act_stress_total}
\end{eqnarray}
%

%
% BibTeX users please use

%\bibliography{mcm.bib,BIO.bib}
%
% Non-BibTeX users please follow the syntax
% the syntax of "referenc.tex" for your own citations

%%%%%%%%%%%%%%%%%%%%%%%% referenc.tex %%%%%%%%%%%%%%%%%%%%%%%%%%%%%%
% sample references
% "physics"
%
% Use this file as a template for your own input.
%
%%%%%%%%%%%%%%%%%%%%%%%% Springer-Verlag %%%%%%%%%%%%%%%%%%%%%%%%%%

%
% BibTeX users please use
% \bibliographystyle{}
% \bibliography{}

%%%%%%%%%%%%%%%%%%%%%%%%%%%%%%%%%%%%%%%%%%%%%%%%%%%%%%%%%%%%%%%%%%%%%%  }

%%%%%%%%%%%%%%%%%%%%%%%%%%%%%%%%%%%%%%%%%%%%%%%%%%%%%%%%%%%%%%%%%%%%%%
% \printindex

% \end{document}

%\include{marchetti/marchetti_chapter}
% \bibliographystyle{pnas}
% \bibliography{lenz/lenz}
%\include{author1}
%\include{author2}

%\backmatter%%%%%%%%%%%%%%%%%%%%%%%%%%%%%%%%%%%%%%%%%%%%%%%%%%%%%%%#
%\appendix
%\include{appendix}
%\printindex

%%%%%%%%%%%%%%%%%%%%%%%%%%%%%%%%%%%%%%%%%%%%%%%%%%%%%%%%%%%%%%%%%%%%%%

\end{document}